\documentclass[apj]{emulateapj}

\usepackage{multirow}
\usepackage{amsmath}
\usepackage{color}

\usepackage{hyperref}	
\hypersetup{colorlinks=true,linkcolor=blue,citecolor=blue,filecolor=blue,urlcolor=blue}

\begin{document}


\title[Uncertainties in Galactic Chemical Evolution Models]
{Uncertainties in Galactic Chemical Evolution Models}

\author{Benoit C\^ot\'e,\altaffilmark{1,2,9,10} Christian Ritter,\altaffilmark{1,9,10} Brian W. O'Shea,\altaffilmark{3,4,9} 
Falk Herwig,\altaffilmark{1,9,10} \\Marco Pignatari,\altaffilmark{5,6,10} Samuel Jones,\altaffilmark{7,10} Chris L. Fryer,\altaffilmark{8,10}}

\altaffiltext{1}{Department of Physics and Astronomy, University of Victoria, Victoria, BC, V8W 2Y2, Canada}
\altaffiltext{2}{National Superconducting Cyclotron Laboratory, Michigan State University, East Lansing, MI, 48824, USA}
\altaffiltext{3}{Department of Physics and Astronomy, Michigan State University, East Lansing, MI, 48824, USA}
\altaffiltext{4}{Department of Computational Mathematics, Science and Engineering, Michigan State University, East Lansing, MI, 48824, USA}
\altaffiltext{5}{E.A. Milne Centre for Astrophysics, Department of Physics \& Mathematics, University of Hull, HU6 7RX, United Kingdom}
\altaffiltext{6}{Konkoly Observatory, Research Centre for Astronomy and Earth Sciences, Hungarian Academy of Sciences, Konkoly Thege Miklos ut 15-17, H-1121 Budapest, Hungary}
\altaffiltext{7}{Heidelberg Institute for Theoretical Studies, Schloss-Wolfsbrunnenweg 35, D-69118 Heidelberg, Germany}
\altaffiltext{8}{Computational Physics and Methods (CCS-2), LANL, Los Alamos, NM, 87545, USA}
\altaffiltext{9}{Joint Institute for Nuclear Astrophysics - Center for the Evolution of the Elements, USA}
\altaffiltext{10}{NuGrid Collaboration, \href{http://nugridstars.org}{http://nugridstars.org}}

\label{firstpage}

\begin{abstract}
We use a simple one-zone galactic chemical evolution model to quantify
the uncertainties generated by the input parameters in numerical
predictions for a galaxy with properties similar to those of the Milky Way.  We
compiled several studies from the literature to gather the current
constraints for our simulations regarding the typical value and uncertainty of the following seven basic
parameters: the lower and upper mass limits of the stellar initial
mass function (IMF), the slope of the high-mass end of the stellar IMF, the
slope of the delay-time distribution function of Type Ia supernovae (SNe Ia), the number of
SNe~Ia per M$_\odot$ formed, the total stellar mass formed, and the
final mass of gas.  We derived a probability
distribution function to express the range of likely values for every parameter, which were then included in
a Monte Carlo code to run several hundred simulations with
randomly selected input parameters.  This approach enables us to
analyze the predicted chemical evolution of 16 elements in a
statistical manner by identifying the most probable solutions along with
their 68\,\% and 95\,\% confidence levels.  Our results show that the
overall uncertainties are shaped by several input parameters that
individually contribute at different metallicities, and thus at
different galactic ages.  The level of uncertainty then depends on the
metallicity and is different from one element to another.  Among the
seven input parameters considered in this work, the slope of the IMF
and the number of SNe~Ia are currently the two main sources of
uncertainty.  The thicknesses of the uncertainty bands bounded by
the 68\,\% and 95\,\% confidence levels are generally within 0.3 and 0.6 dex, respectively.  When
looking at the evolution of individual elements as a function of galactic age
instead of metallicity, those same thicknesses range from 0.1 to 0.6 dex for the 
68\,\% confidence levels and from 0.3 to 1.0 dex for the 95\,\% confidence levels. 
The uncertainty in our chemical evolution model does not include uncertainties relating to stellar yields,
star formation and merger histories, and modeling assumptions.\\
\end{abstract}


\section{Introduction}
\label{sect_intro}
Numerical simulations in astrophysics are challenging because of
their multi-scale nature.   In principle, galactic chemical evolution
models need to resolve both the evolution of stars and the evolution
of galaxies.  In practice, this is problematic because stellar
evolution implies timescales as low as a few seconds in
advanced burning stages, while galaxies have
lifetimes that spread over billions of years.  In addition, stars
occupy an insignificant fraction of the volume of a galaxy, which is a
concern for hydrodynamic simulations.  Of course, it is not possible to
resolve all of these different scales with the computational power
currently available.  In order to consistently follow the chemical
evolution of a galaxy, simulations need to incorporate
\textit{subgrid} treatments to include the physics that cannot be
spatially or temporarily resolved.  This is done by using the
previously calculated outputs of other simulations that fully focus on
the unresolved problem (e.g., stellar evolution), or by using simplified
analytical models.

\begin{figure*}
\begin{center}
\includegraphics[width=6.0in]{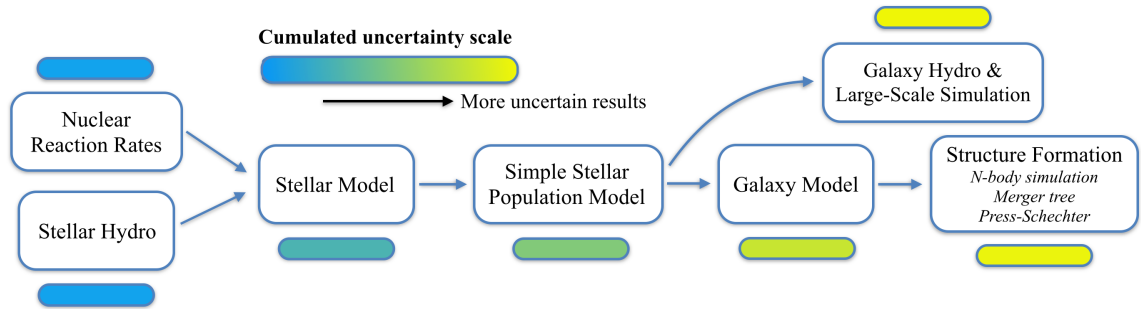}
\caption{Growth of uncertainties in chemical evolution studies from stellar to cosmological scales.  This diagram only serves as a qualitative visualization and may not be complete.}
\label{fig_chain}
\end{center}
\end{figure*}

Stellar models represent the building blocks of chemical evolution
models by providing, in the form of yield tables, the mass ejected by
stars for different elements and metallicities
(e.g., \citealt{pcb98,i99,cl04,t04,mm05,n06,hw10,k10,p13a}).  These stellar
yields are then converted into simple stellar populations (SSPs)
by using stellar lifetimes and an initial mass function (IMF).  
In this context, an SSP refers to a mass element that is 
converted into stars according to a presented star formation rate.
In hydrodynamical simulations, SSP units are used to inject mass locally
around individual \textit{star particles} (e.g., \citealt{od08,w09,hqm12,b14}).  This
procedure can be applied on galactic and cosmological scales.
In semi-analytical models, SSP units are combined together to inject
mass in the gas reservoirs of a galaxy
(e.g., \citealt{bb10,t10,crosby13,y13,delucia14,gomez14,cmd15}).  Usually, these models can be used in a
cosmological context by combining them with N-body simulations and merger trees.

Chemical evolution studies can also be applied to the circumgalactic,
intergalactic, and intracluster medium.  Galactic outflows, gas stripping, and galaxy disruptions are
responsible for entraining enriched material beyond the galactic scale
(see \citealt{mbb12,bl13,s13}).  From the production of the elements
in stars to the enrichment of intergalactic gas, there is a chain of
models integrated with each other.  Stellar models are actually
not the starting point of this chain -- they depend on nuclear physics
for defining the production of new elements in stellar interiors (e.g., \citealt{wkl12}) and on hydrodynamic
experiments and simulations for constraining the behavior of gas
experiencing phenomena such
as turbulence and mixing (e.g., \citealt{ma07,w08,wh15,amy10,h14,sa14})
or exposed to more extreme conditions during supernova explosions
(e.g., \citealt{j12,burrow13,hix14,wmh15}).

Although this chain of models offers a very efficient way to create
solid links between different scales despite the limitation of current
computational resources, it is critical to acknowledge that a great
deal of uncertainty is attached to the numerical predictions coming out of
this chain, particularly at galactic and cosmological scales.   At every scale, each
model deals with their own sets of assumptions and
uncertainties.  Each time a model in the chain uses the results
  of the models that precede it, it ends up implicitly including the
  uncertainties of these preceding models (Figure \ref{fig_chain}).  In that sense, chemical evolution studies 
may already hide potential uncertainties beyond observational errors.
In order to move toward a
better quantification of the global level of uncertainty pertaining to
chemical evolution modeling, it is important to establish the uncertainties at every step along the chain.
In this paper, we focus on the SSP and galactic levels, as a critical component of this bigger picture
between nuclear astrophysics and cosmology simulations.

As clearly demonstrated by \cite{rktm10}, the choice of stellar yields
is a major source of uncertainty in galactic chemical
evolution models (see also \citealt{g97,g02,w09,mcgg15}).  In addition,
all of those yields are affected by nuclear reaction rate uncertainties
(e.g., \citealt{r02,ee04,hal06,tha07,tha09,tha10,p10,p13b,i11,wkl12,pjsr13,t14}) and
by stellar modeling assumptions (e.g., \citealt{whw02,hmm05,k14,kl14,j15,l15}).

Another source of uncertainty is the input parameters associated
with SSPs.  \cite{rcmt05} studied the impact
of the IMF (see also \citealt{mcgg15}) and the stellar lifetimes,
while \cite{msrv09}, \cite{w09}, and \cite{y13} explored the impact of the delay-time
distribution (DTD) function used to calculate the rate of Type~Ia supernovae (SNe Ia).
However, all of these studies focused on the uncertainties resulting from
the different options offered in the literature, and not on the
uncertainties in the measurements that were used to constrain the
input parameters in the first place.

The goal of the present paper is to complement the work
discussed above by using the simplest possible model -- a closed-box,
single-zone model -- and by exploring the impact of the uncertainties
associated with a range of input parameters corresponding to the
measurements of several observationally estimated quantities relating
to the stellar IMF, to SNe Ia properties, and to the global
properties of the Milky Way galaxy.  We do not claim that
a closed-box model is representative of the Milky Way, which is
why we do not compare our results with observational data in this paper.  
We are only using the Galaxy as a test case to constrain the
uncertainty of some of our input parameters in order to understand how
they affect our predictions.  More realistic chemical evolution models
designed to reproduce the Milky Way can be found in the literature
(e.g., \citealt{cmr01,k11,mcm13,mcgg15,shen15,vv15,wpt15}).

Our aim in this work is to place lower bounds on the uncertainties in predictions
relating to some of the most fundamental input parameters in all
chemical evolution models.  In future papers, we will expand
our approach to include galactic inflows and outflows, variable star
formation rates, the effect of mergers and environment, and varied
stellar evolution models and nuclear physics input.
Ultimately, our goal is to obtain a quantification of uncertainty in chemical
evolution modeling.  It will be useful to develop intuition about
what types of predictions from modern chemical evolution models are
reliable when compared
to modern observational data sets.  Furthermore, realistic uncertainties
will be critical when comparing these models to modern
stellar surveys using statistical techniques such as Bayesian Markov Chain
Monte Carlo (e.g., \citealt{gomez12,gomez14,gel14}).  Galactic chemical evolution models are 
an important tool for probing stellar models.  This requires that we know how reliable
the chemical evolution models are.

This paper is organized as follows.  In Section~\ref{sect_code}, we
briefly describe our galactic chemical evolution code and our choice
of stellar yields.  We describe, in Section~\ref{sect_ip}, the seven
input parameters varied in our code and present compilations of the different studies
used to constrain them.  The probability distribution function (PDF), the average value, and the
uncertainty of each parameter are calculated in Section~\ref{sect_pdf}.
We present in Section~\ref{sect_results} the overall uncertainties
in the chemical evolution of 16 elements, along with the individual
contribution of each parameter.  We discuss some caveats and
limitations relating to our results in Section~\ref{sect_disc}. Finally, our conclusions are
given in Section~\ref{sect_conc}.

\section{Chemical Evolution Code}
\label{sect_code}

In this paper, we use the SYGMA module
(C. Ritter et al. in preparation), which stands for Stellar Yields for Galactic
Modeling Applications, to calculate the
composition of the stellar ejecta coming out of SSP units as
a function of time and metallicity using a set of stellar yields.
We also use a simplified version of the OMEGA module (\citealt{cote16}), which
stands for One-zone Model for the Evolution of GAlaxies, to 
combine the contribution of several SSPs to calculate the chemical evolution of a gas reservoir.  These python codes are
part of an upcoming numerical pipeline designed to create permanent
connections between nuclear physics, stellar evolution, and galaxy evolution.  These modules will ultimately be used to probe the
impact of nuclear physics and stellar modeling assumptions on galactic chemical evolution,
but will also provide input data for simulations aiming to study chemical evolution
in a cosmological context.

\subsection{Closed Box Model}
The presented version of OMEGA is a classical one-zone, closed-box galaxy model (see \citealt{ta71}) with a 
continuous star formation history (SFH)
where stars form and inject new elements within
the same gas reservoir, using SYGMA to create an SSP at every timestep.
Single-zone, closed-box models solve the following equation at each
timestep during a simulation (e.g., \citealt{p09}):
\begin{equation}
M_{\rm gas}(t+\Delta t)=M_{\rm gas}(t) + \Big[\dot{M}_{\rm ej}(t) - \dot{M}_\star(t)\Big]\Delta t,
\label{eq_main_sim}
\end{equation}
where, $M_{\rm gas}(t)$ and $\Delta t$ are, respectively, the mass of the gas reservoir at time $t$,
and the duration of the timestep.  $\dot{M}_{\rm ej}(t)$ and $\dot{M}_\star(t)$ are the stellar mass
loss rate and the star formation rate.  In this paper, we ignore
the contribution of galactic inflows and outflows.  As mentioned in Section~\ref{sect_intro}, our goal is
not to present a sophisticated chemical evolution code, but rather to focus on the
impact of basic input parameters on numerical predictions.  

We use a set of stellar yields to calculate the mass ejected by stars (see Section~\ref{sect_stel_yie}), and accordingly the chemical
composition of the gas reservoir is known at any time $t$.  When gas is converted into stars to form an SSP, the mass of each element locked
away is calculated following the current chemical
composition of the gas reservoir.  The metallicity of an SSP formed at time $t$ is then 
represented by the gas metallicity at that time, which is given by
\begin{equation}
Z_\mathrm{gas}(t)=\frac{M_\mathrm{gas}(t) - M_\mathrm{H}(t) -M_\mathrm{He}(t)}{M_\mathrm{gas}(t)}.
\end{equation}
At each timestep, the total stellar
ejecta is calculated by considering the mass $M$, the metallicity
$Z$, and the age $\tau$ of every individual SSP,
\begin{equation}
\dot{M}_{\rm ej}(t)\Delta t=\sum_{i=1}^{N_\mathrm{SSP}}\sum_\mathrm{X=1}^{N_X}\dot{M}_{\mathrm{ej}_X}(M_i,Z_i,\tau_i)\Delta t.
\end{equation}
In this last equation, $i$ and $N_\mathrm{SSP}$ refer to the $i^\mathrm{th}$ SSP and the total number of SSPs
that have formed by time $t$.  $N_X$ represents the number of chemical elements $X$ considered in the stellar yields.
The input parameters used in our model are presented in Section~\ref{sect_ip}.

\subsection{Stellar Yields}
\label{sect_stel_yie}
We use the NuGrid\footnote{\href{http://nugridstars.org}{http://nugridstars.org}} collaboration's yield set 
that includes AGB stars from 1 to $7\,$M$_\odot$ and massive stars from 12 to $25\,$M$_\odot$
at metallicities (given in mass fraction) of $Z=0.02$, $0.01$, $0.006$, $0.001$, and $10^{-4}$ (C. Ritter, private communication).
All data sets are available online\footnote{\href{http://nugridstars.org/projects/stellar-yields}{http://nugridstars.org/projects/stellar-yields}}
and can be explored through WENDI\footnote{\href{http://nugridstars.org/projects/wendi}{http://nugridstars.org/projects/wendi}}.
For this work, we used the yields associated with version 1.0 of the online
NuPyCEE\footnote{\href{http://github.com/NuGrid/NUPYCEE}{http://github.com/NuGrid/NUPYCEE}} (NuGrid Python Chemical Evolution Environment) package.
NuGrid provides all stable elements from hydrogen up to bismuth along with many isotopes.
The complete stellar evolution calculations were performed with MESA (\citealt{p11})
while the post-processing was done using NuGrid's MPPNP code (\citealt{p13a}).
We used the same nuclear reaction rates in all of our calculations.
The explosive nucleosynthesis for massive stars was calculated with the semi-analytical model 
presented in \cite{p13a}.  The original Set1 yields calculated with the GENEC code (\citealt{h04,e08})
for $Z=0.02$ and $Z=0.01$ can be found in \cite{p13a}.  The yields used in this work were
calculated with the same approach and assumptions, but with MESA instead of GENEC.
We complement the presently available NuGrid yields with the SN~Ia yields of \cite{tny86}, which are based on the W7 model of \cite{nt84},
and the zero-metallicity yields of \cite{hw10} who provide masses between 10 and $100\,$M$_\odot$.

\section{Input Parameters}
\label{sect_ip}
In this section, we briefly describe the input parameters used
in our chemical evolution code.  We selected a subset of parameters,
mostly associated with SSPs, to explore the impact of their
uncertainties in our numerical predictions.
For each of these selected parameters, we have gathered a compilation of
observational and numerical studies to constrain their typical value
and uncertainty.  It should be noted that in some studies, the upper
and lower limits of the uncertainty did not have
the same value. However, since we only want to have a general sense of the
current level of uncertainties, we always take the average of the upper
and lower limits in order to work with a single $\pm$ value (as presented
in every table in this section).  For our approach, this approximation
is convenient since we can thereafter apply a Gaussian function on
every considered study to derive the PDF
of our input parameters (see Section \ref{sect_pdf}).

\subsection{Lower Mass Limit of the IMF}
The minimum stellar mass, often called the hydrogen-burning minimum
mass, refers to the lowest possible mass for an object to ignite
nuclear fusion (see \citealt{cb00,k13}).  Although the lower mass
limit of the IMF, $M_\mathrm{low}$, can be measured observationally
(\citealt{b13}), the generally adopted value for the minimum stellar
mass comes from the predictions of evolutionary stellar models, which
can be seen in Table \ref{tab_uncer_m_low}.  Most of the uncertainties
contained in this last table represent the mass separation between
models producing brown dwarfs and models producing stars.  The
resulting uncertainties should then be considered as lower limits,
since modeling assumptions must add a significant (albeit difficult
to quantify) degree of uncertainty.

\begin{table}
\caption{Compilation of the Predicted Minimum Stellar Mass.}
\centering
\begin{tabular}{cc}
\noalign{\medskip}
\hline
\hline
\noalign{\medskip}
References & $M_{\rm min}$ [M$_\odot$] \\
\noalign{\medskip}
\hline
\noalign{\medskip}
Hayashi \& Nakano~(\citeyear{hn63}) & 0.075 $\pm$ 0.005 \\
\noalign{\medskip}
Kumar~(\citeyear{k63}) & 0.080 $\pm$ 0.010 \\
\noalign{\medskip}
Grossman \& Graboske~(\citeyear{gg71}) & 0.085 $\pm$ 0.005 \\
\noalign{\medskip}
Straka~(\citeyear{s71}) & 0.085 $\pm$ 0.004 \\
\noalign{\medskip}
Burrows et~al.~(\citeyear{b97}) & 0.078 $\pm$ 0.0025 \\
\noalign{\medskip}
Chabrier \& Baraffe~(\citeyear{cb00}) & 0.079 $\pm$ 0.004 \\
\noalign{\medskip}
\hline
\label{tab_uncer_m_low}
\end{tabular}
\end{table}

As long as $M_\mathrm{low}$ does not change within an order of magnitude,
its exact value is not crucial for SSPs and galactic chemical evolution, since
these stars have  lifetimes that are too long to contribute to the
chemical enrichment process.  However, modifying the lower mass limit of the IMF changes the number
of more massive stars in stellar populations (by changing the fractional amount of massive stars), which can in turn modify the rate with which the galactic gas
gets enriched.

\subsection{Upper Mass Limit of the IMF}
\label{sect_m_up}
As with the lower mass limit, the upper mass limit $M_{\rm up}$ of the IMF has
an impact on the total number of stars that participate in the
chemical enrichment process.  This parameter can be estimated by looking at the most massive
component of stellar clusters.  To do so, observations need to focus
on clusters young enough (typically with ages less than $\sim$ 3Myr) so that
the current stellar mass function is as close as possible to the
actual IMF.  Moreover, the clusters need to be massive enough to allow
a comprehensive sampling of the IMF at the high-mass end.  According to observations, the maximum stellar
mass observed in a cluster seems to reach a ceiling value when the
cluster has a total stellar mass above roughly 10$^4\,$M$_\odot$ (\citealt{wk06,wkb10,wkp13}).  We present in Table
\ref{tab_uncer_m_up} a compilation of the mass of the most massive
star observed in stellar clusters that respect the above conditions.
It is worth noting that the derived stellar mass depends on the
stellar model used to match observations (\citealt{m15}).

\begin{table}
\caption{Compilation of the maximum stellar mass observed in stellar clusters.
From left to right, the columns represent the reference paper, the observed
  stellar cluster, and the mass of the most
  massive star with its uncertainty.}
\centering
\begin{tabular}{ccc}
\noalign{\medskip}
\hline
\hline
\noalign{\medskip}
References & Stellar Cluster & $M_{\rm up}$ [M$_\odot$] \\
\noalign{\medskip}
\hline
\noalign{\medskip}
Crowther et al.~(\citeyear{c10}) & \multirow{3}{*}{NGC 3603} & 166 $\pm$ 20 \\
Weidner et al.~(\citeyear{wkb10})$^{*}$ & & 150 $\pm$ 50 \\
Weidner et al.~(\citeyear{wkp13}) & & 121 $\pm$ 35 \\
 \noalign{\medskip}
Weidner \& Kroupa~(\citeyear{wk06}) & \multirow{3}{*}{Trumpler 14/16} & 120 $\pm$ 15 \\
Weidner et al.~(\citeyear{wkb10})$^{*}$ &  & 150 $\pm$ 50 \\
Weidner et al.~(\citeyear{wkp13}) &  & 100 $\pm$ 45 \\
  \noalign{\medskip}
Figer~(\citeyear{f05}) & \multirow{4}{*}{Arches} & 126 $\pm$ 15 \\
Crowther et al.~(\citeyear{c10}) &  & 135 $\pm$ 15 \\
Weidner et al.~(\citeyear{wkb10})$^{*}$ &  & 135 $\pm$ 15 \\
Weidner et al.~(\citeyear{wkp13}) &  & 111 $\pm$ 40 \\
\noalign{\medskip}
Wu et al.~(\citeyear{w14}) & W49 & 140 $\pm$ 40 \\
 \noalign{\medskip}
Oey \& Clarke~(\citeyear{oc05}) & 9 clusters & 160 $\pm$ 40 \\
\noalign{\medskip}
\hline
\multicolumn{3}{l}{$^*$\footnotesize{ We took the value that was not present in the compilation}} \\
\multicolumn{3}{l}{\footnotesize{ \,\,\,of Weidner et~al.~(\citeyear{wkp13}).}} \\
\label{tab_uncer_m_up}
\end{tabular}
\end{table}

The stellar cluster R136, which hosts a very massive star with a
possible initial mass around $320\,$M$_\odot$ (\citealt{c10}), is not
included in our compilation.  This extreme case points toward a higher
upper mass limit than the ones shown in Table~\ref{tab_uncer_m_up} (see also \citealt{ph14}).  The possible observation of a
pair-instability SN presented in \cite{gy07} also supports the
existence of very massive stars in the nearby Universe.  However, such
massive stars could be the product of binary interactions and stellar
mergers, and not the result of the star formation process
(\citealt{bko12,plk12,fpz13,setal14}).  For this reason, we choose to
exclude R136 from our sample.  In addition, there is a general
agreement that the upper mass of the IMF should be around
$150\,$M$_\odot$ (\citealt{wk04,f05,oc05,k06,zy07,k13}).

This implies, however, that we are
excluding pair-instability SNe in our calculations, although they could
be important at low metallicity (e.g., \citealt{hw02,cm14,kyl14}).
We also ignore, for now, the contribution of hypernovae.  We refer to \cite{n13} and \cite{kobay06,kobay14}
for more information on the impact of those high-energy explosions in the chemical 
evolution of galaxies.  As explained below in Section~\ref{sect_y_uml}, we do not
consider stellar yields for stars more massive than 30\,M$_\odot$.  Therefore,
the upper mass limit of the IMF affects the total mass of gas locked
into stars, but not the chemical evolution.  The uncertainty caused by $M_{\rm up}$
is therefore significantly underestimated in our work.  Nevertheless, we decided to
present our compilation (Table~\ref{tab_uncer_m_up}) for the sake of completeness
and for future reference.

\subsection{Slope of the IMF}
The IMF is certainly the most important aspect to consider when
modeling a stellar population, because it sets the number ratio of
low-mass to massive stars.  In the original IMF proposed by \cite{s55},
$\xi(M_i)\propto M_i^{-\alpha}$, a unique $\alpha$ index of 2.35
was used for the entire stellar mass range.  However, more recently, for $M_i$ $<$ $1\,$M$_\odot$, the IMF has been
modified to account for the lower number of observed stars compared to
the Salpeter IMF (e.g., \citealt{k01,c03}).
Through their winds and explosions, stars with different initial
masses $M_i$ do not eject the same type or relative quantities of elements into their
surroundings.  Varying the IMF can therefore have an impact
on the predicted chemical evolution of a galaxy.  Although many
studies support the idea of a universal IMF (see \citealt{bcm10}),
there are still uncertainties associated with the exact value of its
slope at the high-mass end.

\begin{table}
\caption{Compilation of estimates of the slope of the high-mass end of the observed
  stellar initial mass function.  From left
  to right, the columns represent the reference paper, the target
  galaxy used to derived the IMF, and the slope of the IMF with its
  uncertainty.  LMC and SMC stand for Large and Small
  Magellanic Cloud.}
\centering
\begin{tabular}{ccc}
\noalign{\medskip}
\hline
\hline
\noalign{\medskip}
References & Galaxy & $\alpha$ \\
\noalign{\medskip}
\hline
\noalign{\medskip}
Salpeter (\citeyear{s55}) & Milky Way & 2.35 $\pm$ 0.20 \\
\noalign{\medskip}
\multirow{3}{*}{Massey (\citeyear{m98})} & Milky Way & 2.26 $\pm$ 0.34 \\ 
 & LMC & 2.37 $\pm$ 0.26 \\
 & SMC & 2.30 $\pm$ 0.10 \\
\noalign{\medskip}
Massey \& Hunter(\citeyear{mh98}) & LMC & 2.30 $\pm$ 0.10 \\
\noalign{\medskip}
Kroupa (\citeyear{k01}) & Milky Way & 2.30 $\pm$ 0.70 \\
\noalign{\medskip}
Slesnick et al. (\citeyear{shm02}) & Milky Way & 2.30 $\pm$ 0.20 \\
\noalign{\medskip}
Baldry \& Glazebrook (\citeyear{bg03}) & Luminosity$^*$ & 2.15 $\pm$ 0.20 \\
\noalign{\medskip}
Chabrier (\citeyear{c03}) & Milky Way & 2.30 $\pm$ 0.30 \\
\noalign{\medskip}
Dib (\citeyear{d14}) & Milky Way & 2.07 $\pm$ 0.25 \\
\noalign{\medskip}
\multirow{3}{*}{Weisz et al. (\citeyear{w15})} &  M31 & 2.45 $\pm$ 0.045 \\
 & Milky Way & 2.16 $\pm$ 0.10 \\
 & LMC & 2.29 $\pm$ 0.10 \\
\noalign{\medskip}
\hline
\multicolumn{3}{l}{$^*$\footnotesize{ Derived from galaxy luminosity densities.}} \\
\label{tab_uncer_imf}
\end{tabular}
\end{table}

In the present work, we use the IMF of \cite{c03}, where we modify the
slope of the power law for stars more massive than $1\,$M$_\odot$.  This
approach was proposed by \cite{w15} who observed a steep slope of 2.45
in the IMF of the M31 galaxy.  We use the compilation
presented in Table \ref{tab_uncer_imf} to set the slope of the IMF, however.
Aside from the work of \cite{s55}, \cite{k01}, and \cite{c03}, every
study presented in Table~\ref{tab_uncer_imf} focuses on the IMF of massive stars
by looking at young stellar clusters.

\subsection{Delay-Time Distribution of SNe Ia}
As opposed to core-collapse supernovae (CC SNe) that always explode at the end of the lifetime of
massive stars (e.g., \citealt{z38,b90,h03}), a certain time is needed for SNe Ia to occur after the
emergence of white dwarfs (e.g., \citealt{hkrr13}).  Their rate of appearance must then be
modeled by considering a delay-time distribution (DTD) function,
$\phi(t)$, that can be seen as the probability of a white dwarf to explode
as a function of time.  For any SSP, following
\cite{g05} and \cite{w09}, we define the rate of explosion as

\begin{equation}
\label{eq_R_Ia}
R_\mathrm{Ia}(t)=A_\mathrm{Ia}f_\mathrm{wd}(t)\phi(t),
\end{equation}

\noindent where $A_\mathrm{Ia}$ and $f_\mathrm{wd}(t)$ represent a
normalization constant and the fraction of progenitor stars that have
turned into white dwarfs by time $t$, which are stars with initial
mass between 3 and $8\,$M$_\odot$ (e.g., \citealt{mm12}).  The fraction
of white dwarfs is calculated from the lifetime of
intermediate-mass stars and the IMF.  This quantity thus depends on
the input parameters that characterize the slope of the IMF and its mass boundary.
For the DTD function, we adopt a power law
of the form $t^{-\beta}$ with $\beta$ being close to unity
(\citealt{mmn14}).  This function implies a prompt appearance of
explosions and is in good agreement with the higher rate of SNe Ia
observed in bluer galaxies (\citealt{m05,l11}).  

\begin{table}
\caption{Compilation of the observed slope associated with the delay-time distribution function of
  SNe~Ia.  From left to right, the columns represent the reference
  paper, the indicator used to derive the DTD function along with the SN Ia
  rate, and the slope of the DTD function with its uncertainty.  All
  references in this table have been taken from the review of
  Maoz~et~al.~(\protect\citeyear{mmn14}).}
\centering
\begin{tabular}{ccc}
\noalign{\medskip}
\hline
\hline
\noalign{\medskip}
References & Indicators & $\beta$  \\
\noalign{\medskip}
\hline
\noalign{\medskip}
Totani et~al.~(\citeyear{t08}) & Elliptical galaxies & 1.08 $\pm$ 0.15 \\
\noalign{\medskip}
Maoz et~al.~(\citeyear{msg10}) & Galaxy clusters & 1.20 $\pm$ 0.30 \\
\noalign{\medskip}
Graur et~al.~(\citeyear{g11}) & Cosmic SFH & 1.10 $\pm$ 0.27 \\
\noalign{\medskip}
Maoz et~al.~(\citeyear{mmb12}) & Individual galaxies & 1.07 $\pm$ 0.07 \\
\noalign{\medskip}
Perrett et~al.~(\citeyear{p12}) & Cosmic SFH & 1.07 $\pm$ 0.15 \\
\noalign{\medskip}
Graur et~al.~(\citeyear{g14}) & Cosmic SFH & 1.00 $\pm$ 0.16 \\
\noalign{\medskip}
\hline
\label{tab_uncer_dtd}
\end{tabular}
\end{table}

Measuring $\beta$ is not an easy task since it relies on the
observation of only a few SNe Ia distributed across cosmic time.  In
general, the main idea is to connect the evolution of the explosion
rate to a quantity that probes the formation of the progenitor stars,
which is often the cosmic SFH or the SFH of individual galaxies.
Table \ref{tab_uncer_dtd} presents a compilation of different studies
that derived the slope of this DTD function.  We refer to \cite{mmn14}
for a review of the different methods commonly used to derive the
shape of this function, and to \cite{mr01}, \cite{s04}, \cite{matt06}, \cite{phs08}, and
\cite{rbf09,r11} for alternative forms of DTD functions.

\begin{table}
\caption{Compilation of the derived normalization for the number of SNe~Ia formed in a
  stellar population.  From left to right, the columns represent the
  reference paper, the number of observed SNe Ia, and the derived
  number of SNe Ia per stellar mass formed with its uncertainty.  For reference, the number of CC SNe in a stellar
  population is about 10$^{-2}$ SN M$_\odot^{-1}$ for a Chabrier IMF.}
\centering
\begin{tabular}{ccc}
\noalign{\medskip}
\hline
\hline
\noalign{\medskip}
\multirow{2}{*}{References} & \multirow{2}{*}{Nb SNe Ia} & $N_{\rm Ia}$ \\
 & & [$10^{-3}$ SNe M$_\odot^{-1}$] \\
\noalign{\medskip}
\hline
\noalign{\medskip}
Maoz et~al.~(\citeyear{msg10}) & SN rates$^*$ & 4.65 $\pm$ 1.25 \\
\noalign{\medskip}
Graur et~al.~(\citeyear{g11}) & 96 & 1.00 $\pm$ 0.50 \\
\noalign{\medskip}
Maoz et~al.~(\citeyear{m11}) & 82 & 2.30 $\pm$ 0.60 \\
\noalign{\medskip}
Maoz \& Mannucci~(\citeyear{mm12}) & SN rates  & 2.00 $\pm$ 1.00 \\
\noalign{\medskip}
Maoz et~al.~(\citeyear{mmb12}) & 132 & 1.30 $\pm$ 0.15 \\
\noalign{\medskip}
Perrett et~al.~(\citeyear{p12}) & 691 & 0.57 $\pm$ 0.15 \\
\noalign{\medskip}
Graur \& Maoz~(\citeyear{gm13}) & 90 & 0.80 $\pm$ 0.40 \\
\noalign{\medskip}
Graur et~al.~(\citeyear{g14}) & 13 & 0.90 $\pm$ 0.40 \\
\noalign{\medskip}
Rodney et~al.~(\citeyear{r14}) & 24 & 0.98 $\pm$ 1.03 \\
\noalign{\medskip}
\hline
\multicolumn{3}{l}{$^{*}$\footnotesize{The authors used SN rate compilations to derive $N_\mathrm{IA}$.}} \\
\label{tab_uncer_NIa}
\end{tabular}
\end{table}

\subsection{Number of SNe Ia}
The normalization of the rate of SNe Ia is an important parameter
since it sets the total number of Type Ia explosions occurring during
the lifetime of a galaxy.  For any SSP formed
during a simulation, the normalization constant can be calculated by
integrating equation~(\ref{eq_R_Ia}),

\begin{equation}
A_{\mathrm{Ia}}=\frac{N_\mathrm{Ia}M_{\mathrm{SSP}}}{\int_0^{t_H}f_\mathrm{wd}(t)\phi(t)dt},
\label{eq_A_Ia}
\end{equation}
\noindent where $N_\mathrm{Ia}$, $M_{\mathrm{SSP}}$, and $t_H$ are,
respectively, the total number of SNe Ia per unit of stellar mass
formed, the mass of the SSP, and the Hubble time.  In that last equation,
every quantity is known besides $N_\mathrm{Ia}$. We assume that an SSP is 
created at each timestep, and calculate its corresponding mass, $M_{\mathrm{SSP}}$,
from the star formation rate and the duration of the timestep at the time of formation.

In the case of CC SNe, the number of explosions per stellar mass
formed is easily calculable from the IMF (but see Section \ref{sect_y_uml}) and one could in principle
use the rate ratio of CC SNe relative to SNe Ia observed in the local
Universe to derive $N_\mathrm{IA}$.  However, only looking at current
rates is misleading because SNe Ia may come from any old or young
stellar population, whereas CC SNe only probe stellar populations
younger than about 40 Myr.  This ratio is far from universal: SNe Ia are more
frequent than CC SNe in elliptical galaxies, and CC SNe are more
frequent than SNe Ia in star-forming spiral galaxies
(\citealt{c97,c99,m05,l11}).  

Therefore, in order to constrain the normalization constant of SNe Ia
for individual SSPs, which should not depend on the type of galaxy,
we need to rely on works that integrated the explosion rates over a
significant amount of time, ideally across the Hubble time.  Because
the rate of SNe Ia is usually given in units of [SN yr$^{-1}$
M$_\odot^{-1}$], its integration over time yields a value in units of
[SN M$_\odot^{-1}$] that can directly be used in equation
(\ref{eq_A_Ia}) for any stellar population.  Table \ref{tab_uncer_NIa}
presents different studies that provide such a value.  We again refer
to \cite{mmn14} for more information about the methodology behind the
normalization of DTD functions.

\subsection{Current Stellar Mass}
\label{sect_cur_stel_mass}
To normalize our SFH (see Section~\ref{sect_sfh}), we use the current stellar mass, $M_{\star,f}$,
of the Milky Way (Table~\ref{tab_uncer_M_stel}).  This quantity can
be defined by
\begin{equation}
M_{\star,f}=M_{\star,\mathrm{tot}} - M_\mathrm{ej},
\end{equation}
where $M_{\star,\mathrm{tot}}$ and $M_\mathrm{ej}$ represent, respectively,
the total stellar mass obtained by integrating the SFH and the total mass of gas returned by all SSPs.
Before running a simulation, we calculate $f_\mathrm{ej}$, which is the average fraction of gas returned by SSPs, using
the IMF and our five non-zero-metallicity sets of yields.  The total stellar mass formed
in our simulations is then given by
\begin{equation}
M_{\star,\mathrm{tot}}=\frac{M_{\star,f}}{1-f_\mathrm{ej}}.
\end{equation}
We found that the fraction of
gas returned by our SSPs ranges from 0.25 to 0.51 with an 
average value of 0.36.  However, we did not include ejecta for stars
more massive than 30\,M$_\odot$ (see Section~\ref{sect_y_uml}).

\begin{table}
\caption{Compilation of the derived current stellar mass of the Milky
  Way. The contribution of halo stars is not included in this
  table.}
\centering
\begin{tabular}{ccc}
\noalign{\medskip}
\hline
\hline
\noalign{\medskip}
References & Methodology & $M_{\star}$ [$10^{10}\,$M$_\odot$]\\
\noalign{\medskip}
\hline
\noalign{\medskip}
Flynn et~al.~(\citeyear{f06}) & Luminosity density & 5.15 $\pm$ 0.80 \\
\noalign{\medskip}
McMillan~(\citeyear{mcm11}) & Bayesian analysis & 6.43 $\pm$ 0.63 \\
\noalign{\medskip}
Bovy \& Rix~(\citeyear{bv13}) & Stellar dynamics & 5.20 $\pm$ 1.70 \\
\noalign{\medskip}
Licquia \& & \multirow{2}{*}{Bayesian analysis} & \multirow{2}{*}{6.08 $\pm$ 1.14} \\
Newman~(\citeyear{ln14}) & & \\
\noalign{\medskip}
\hline
\label{tab_uncer_M_stel}
\end{tabular}
\end{table}

\subsection{Current Mass of Gas}
\label{sect_cgf}
The last parameter considered in this work is the final mass of gas, $M_{\mathrm{gas},f}$,
present at the end of our simulations.  According to \cite{kpa15}, the current mass of gas
in the Milky Way is $(8.1\pm4.5)\times10^{9}$\,M$_\odot$ for the disk, and $(1.1\pm0.8)\times10^{9}$\,M$_\odot$ for
the bulge.  Because we use a single-zone model, we combine these two values to
obtain $M_{\mathrm{gas},f}=(9.2\pm5.3)\times10^{9}$\,M$_\odot$.
We use $M_{\mathrm{gas},f}$ to calculate the initial mass of gas, $M_{\mathrm{gas},i}$, present 
at the beginning of our simulations.  Since we use a closed-box model, this quantity is simply defined by
\begin{equation}
\label{eq_gas_f_i}
M_{\mathrm{gas},i}=M_{\mathrm{gas},f}+M_{\star,f}.
\end{equation}

\subsection{Star Formation History}
\label{sect_sfh}
To derive the SFH of our simulated galaxies, we use the following relation between the star formation rate and the mass of gas present at time $t$ (e.g., \citealt{sd14}),
\begin{equation}
\label{eq_rec_eq_0}
\dot{M}_\star(t) = f_\star M_\mathrm{gas}(t),
\end{equation}
where $f_\star$ is a constant free parameter and represents the star formation efficiency in units of [yr$^{-1}$].
Before running a simulation, we use
a recurrence formula derived from equation~(\ref{eq_main_sim}) to find the value of $f_\star$ that will
generate the right gas-to-stellar mass ratio at the end of our simulations.  Using the $f_\mathrm{ej}$
parameter defined in Section~\ref{sect_cur_stel_mass} to represent the mass returned by stars,
the approximated evolution of the gas reservoir for the $n^{\mathrm{th}}$ step is given by
\begin{equation}
\label{eq_rec_eq_1}
M_\mathrm{gas}(t_{n+1})\approx M_\mathrm{gas}(t_n) + f_\mathrm{ej}\dot{M}_\star(t_n)\Delta t_n - \dot{M}_\star(t_n)\Delta t_n.
\end{equation}
Substituting equation~(\ref{eq_rec_eq_0}) into equation~(\ref{eq_rec_eq_1}) yields 
\begin{equation}
M_\mathrm{gas}(t_{n+1})\approx M_\mathrm{gas}(t_n) + f_\star\left(f_\mathrm{ej}-1\right)M_\mathrm{gas}(t_n)\Delta t_n,
\end{equation}
which simplifies to
\begin{equation}
\label{eq_rec_eq_2}
M_\mathrm{gas}(t_{n+1})\approx M_\mathrm{gas}(t_n)\left[1+f_\star\left(f_\mathrm{ej}-1\right)\Delta t_n\right].
\end{equation}
Starting with $n=0$, $M_\mathrm{gas}(t_0)=M_{\mathrm{gas},i}$, and a first guess for $f_\star$,
equation~(\ref{eq_rec_eq_2}) is looped over all the pre-defined timesteps of the forthcoming simulation to calculate
the final mass of gas.  If the final gas content differs by more than 1\,\% of the desired value, which is $M_{\mathrm{gas},f}$
(see Section~\ref{sect_cgf}), the operation is repeated with a revised value for $f_\star$
until the criteria is respected.  By design, since the initial mass of gas is $M_{\mathrm{gas},i}$
and the final mass of gas is $\sim$\,$M_{\mathrm{gas},f}$, the selected star formation efficiency
will form the right amount of stars (see equation~\ref{eq_gas_f_i}).

We found that our approximation
only deviates by less than 2\,\% from the actual values recovered at the end of our simulations.  The
SFH is calculated with the $M_{\star,f}$ and $M_{\mathrm{gas},f}$ parameters and is therefore affected
by their uncertainties.  Because the $f_{\mathrm{ej}}$ parameter is implied in the calculation, the
derived SFH also depends on the IMF.

\subsection{Other Parameters}
\label{sect_other_param}
Besides the seven input parameters described above, there
are other parameters used in the calculation that we do not include in our uncertainty calculation.
Those additional parameters, described in the next sub-sections, do not
necessarily have direct observational constraints with measurable uncertainties.
Furthermore, some of those parameters are more associated with
modeling assumptions than with observable quantities.
Because of the lack of quantified uncertainties, it is difficult
to include them in our Monte Carlo calculation (see Section \ref{sect_pdf}).
As a result, we simply decided to ignore their impact in this analysis. 
\\
\subsubsection{Upper Mass Limit for CC~SNe Progenitors}
\label{sect_y_uml}
As described in Section \ref{sect_m_up}, it is possible for molecular clouds to
form very massive stars up to $\sim$~$100\,$M$_\odot$.  However, the
fate of such massive stars is still uncertain and not fully understood.  
In addition, many CC~SNe yields do not provide progenitor stars
more massive than $40\,$M$_\odot$ (e.g., \citealt{ww95,
cl04,n06}), although yields for pair-instability SNe are available (e.g., \citealt{kyl14}).  In order to cover the entire stellar mass range in chemical evolution
models, the yields of the most massive stars can be extended and used to
represent all of the more massive stars included in the mass range covered by the IMF.  However, doing so implies that
all massive stars produce a CC~SN at the end of their lifetime, which is not in
agreement with the black hole mass distribution observed in our Galaxy (\citealt{bbf12,fbw12}).
In fact, numerical studies have shown that many massive stars should directly collapse
into black holes instead of producing an explosion (e.g., \citealt{zwh08,u12}).  According to \cite{whw02} and \cite{h03},
this is most likely to be the case for stars more massive than $25\,$M$_\odot$.

\begin{figure*}
\includegraphics[width=7in]{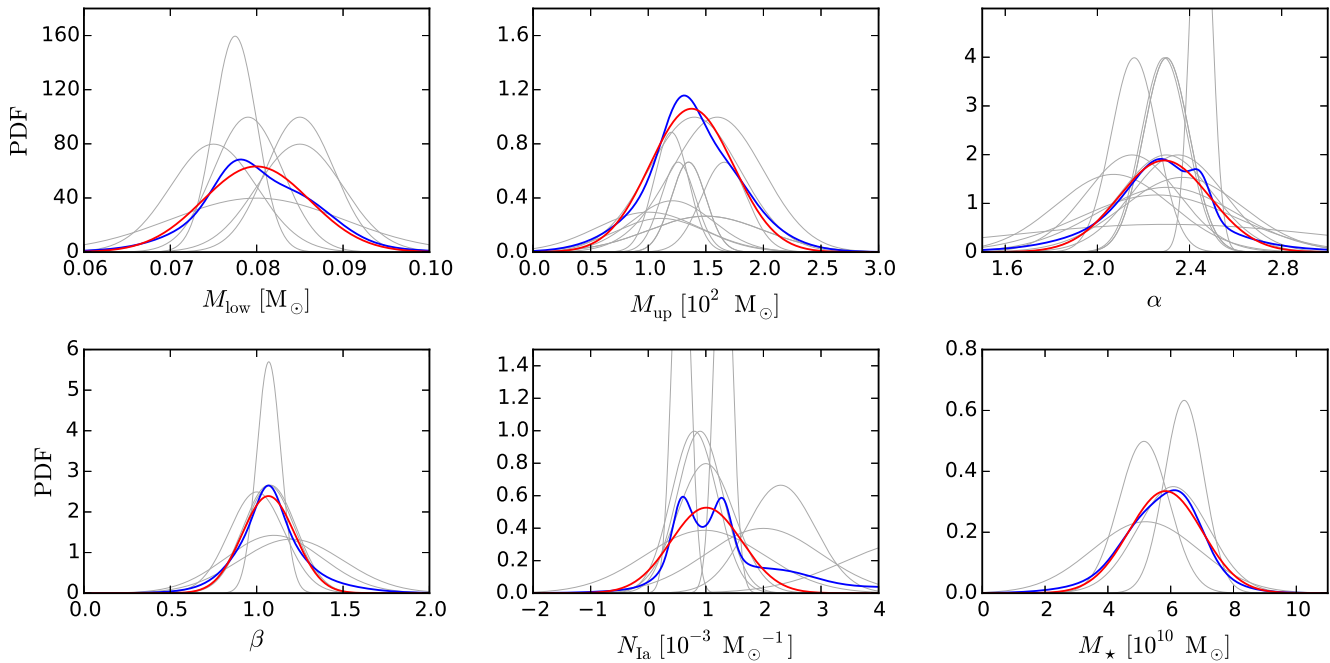}
\caption{Probability distribution functions of six of our seven input
  parameters.  The gray, blue, and red lines represent respectively
  the different observational constraints, the normalized sum of these
  constraints, and the Gaussian fit of the normalized sum.  In the $M_\mathrm{up}$ panel,
  most of the gray lines are actually not normalized to
  one.  Since the mass of the most massive star varies from one stellar
  cluster to another, we wanted each cluster considered in Table
  \ref{tab_uncer_m_up} to have the same statistical weight.  To do so,
  the normalization factor of each study has been divided by the number
  of studies that focused on the same cluster.}
\label{fig_param}
\end{figure*}

Because the most massive stars available in our set of yields is $25\,$M$_\odot\,$(except
for zero-metallicity stars), and because of the observed black hole mass distribution,
we introduce an initial stellar mass threshold $M_\mathrm{thresh}$ of $30\,$M$_\odot$ above which stars do not release any
ejecta.  We added this parameter by simplicity, since we are not sure yet how to
treat the most massive stars within a SSP.  It is beyond the scope of this paper to study how
this threshold impacts our results.  However, because of this choice, the impact
of the upper mass limit of the IMF is underestimated in our work (see Section~\ref{sect_m_up}).
It is worth noting that an initial stellar mass threshold may not exist.  Recent
simulations by \cite{u12} suggests that direct black hole formation and successful CC explosion
are both possible outcomes for progenitor stars more massive than 15~M$_\odot$, with no
direct correlation with the initial stellar mass.  A further source of complications is given
by the lack of observations of CC~SN progenitors with initial mass above $\sim$~$18\,$M$_\odot$ (\citealt{smartt15}), 
which is far below what is usually assumed in chemical evolution models.

\subsubsection{Transition Metallicity}
\label{sect_trans_Z}
All of our simulations start with a gas reservoir that has a primordial composition.
We first use the zero-metallicity yields of \cite{hw10} until the gas reservoir becomes enriched
with metals.  When that happens, we switch and instead use the yields for $Z=10^{-4}$ (currently the lowest
metallicity available with NuGrid), until the metallicity of the gas actually reaches $Z=10^{-4}$ ([Fe/H]~$\simeq$~$-2.3$), above which
we start to interpolate the yields according to the metallicity.  We therefore
never interpolate the yields when the gas reservoir has a non-zero metallicity between $Z=0$ and $Z=10^{-4}$,
since we use the logarithm of $Z$ for our interpolations.
There is, instead, a transition metallicity
$Z_\mathrm{trans}$ above which we stop using the zero-metallicity yields. 
It is worth noting that several studies do support the existence of a transition
metallicity around 10$^{-5}-10^{-3}$~$Z_\odot$, between metal-free and metal-poor stars
(\citealt{b01,bromm03,s06,cgk08,d11,s12,mso14,s15}).  However, in our case, we do not
fix the value of $Z_\mathrm{trans}$.  It simply refers to the first non-zero metallicity occurring in the simulations, which can differ from
one run to another depending on the values of our input parameters.

\subsubsection{Minimum Mass for CC~SNe}
In the calculation of the stellar ejecta coming out of SSPs, we assume a transition initial
stellar mass $M_\mathrm{trans}$, fixed a $8\,$M$_\odot$, above which stars produce CC~SNe.
According to several studies, this delimitation between intermediate-mass and massive stars
should roughly be between 7 and $10\,$M$_\odot$ (e.g., \citealt{t96,p08,s09,j13,f15,wheg15}). The value of this
parameter is mainly affected by the physics assumptions made in stellar models.  Since our work
in this paper focusses on the impact of uncertainties in the measurements of input parameters,
and not on the impact of stellar modeling assumptions, $M_\mathrm{trans}$ is kept
constant at $8\,$M$_\odot$ in all of our simulations, although recent 
simulations of super-AGB stars suggest a higher value (see \citealt{p08,j13,f15}).
We do, however, consider the lower mass limit
of the IMF in the uncertainty calculation even if its value and uncertainty depends on modeling assumptions
in stellar models.  We only do this because we want to have a complete sampling of the parameters
describing the IMF.

\subsubsection{SNe Ia Progenitors}
In the calculation of the rate of SNe Ia and its normalization, we need
to define the explosion progenitors in order to calculate the fraction of white dwarfs
used in equations (\ref{eq_R_Ia}) and (\ref{eq_A_Ia}).  According to many 
studies, those progenitors should be stars with initial masses between
3 and $8\,$M$_\odot$ (e.g., \citealt{d04,mm12}).  However, there is no uncertainty derived for the
lower and upper limits of this mass interval.  Using 3 and $8\,$M$_\odot$ is
therefore an assumption rather than a measurement with an uncertainty.
We do not consider the different evolutionary channels and the possibility that
different types of SN~Ia may contribute to the chemical inventory of galaxies with relatively different
delay times and different numbers (e.g., \citealt{hkrr13,setal13,kobay15,metal15}).
Therefore, the SN~Ia contribution must be considered in our work to be an average
representative of different potential SN~Ia populations.

\section{Probability Distribution Functions}
\label{sect_pdf}
In order to consider all the studies compiled in the previous section,
we first associated a Gaussian function to
each one of them, by centering the function on the mean value, 
taking the uncertainty as the standard deviation $\sigma$, and finally
normalizing to one (gray lines in Figure
\ref{fig_param}).  Our goal here is not to investigate which works are
the most relevant, but rather to use all of them to establish an order of magnitude of
the current state of uncertainties regarding these observations.
For each parameter, we therefore
summed all the Gaussians, normalized the resulting curve to one (blue
lines), and fitted a new Gaussian function on top of this resulting
curve (red lines).

\begin{table*}
\caption{List of the seven input parameters used in this work 
    to generate uncertainties in our numerical predictions.
    The value and uncertainty of each parameter are 
    the mean value and the standard deviation taken from the 
    corresponding Gaussian fit presented in Figure \protect\ref{fig_param}.}
\centering
\begin{tabular}{cccc}
\noalign{\medskip}
\hline
\hline
\noalign{\medskip}
Parameter & Description & Typical value & References \\
\noalign{\medskip}
\hline
\noalign{\medskip}
\multicolumn{4}{c}{\bf{Simple Stellar Population}} \\
\noalign{\medskip}
\noalign{\medskip}
$M_{\rm low}$ & Lower mass limit of the IMF [$10^{-2}\,$M$_\odot$] & 8.00 $\pm$ 0.62 & Table \ref{tab_uncer_m_low} \\
\noalign{\medskip}
$M_{\rm up}$ & Upper mass limit of the IMF [M$_\odot$] & 138 $\pm$ 36 & Table \ref{tab_uncer_m_up} \\
\noalign{\medskip}
$\alpha$ & Slope of the IMF & 2.29 $\pm$ 0.20 & Table \ref{tab_uncer_imf} \\
\noalign{\medskip}
$\beta$ & Slope of the DTD of SNe Ia & 1.07 $\pm$ 0.15 & Table \ref{tab_uncer_dtd} \\
\noalign{\medskip}
$N_{\rm Ia}$ & Number of SNe Ia [$10^{-3}$ M$_\odot^{-1}$] & 1.01 $\pm$ 0.62 & Table \ref{tab_uncer_NIa} \\
\noalign{\medskip}
\hline
\noalign{\medskip}
\multicolumn{4}{c}{\bf{Galaxy}} \\
\noalign{\medskip}
\noalign{\medskip}
$M_\star$ & Current stellar mass$^{*}$ [$10^{10}\,$M$_\odot$] & 5.84 $\pm$ 1.17 & Table \ref{tab_uncer_M_stel} \\
\noalign{\medskip}
$M_{\mathrm{gas},f}$ & Current mass of gas$^{**}$ [$10^{9}\,$M$_\odot$] & 9.2 $\pm$ 5.3 & \cite{kpa15} \\
\noalign{\medskip}
\hline
\multicolumn{4}{l}{$^{*}$\,\,\,\footnotesize{This parameter is used to calibrate the SFH (see Section~\ref{sect_cur_stel_mass}).}} \\
\multicolumn{4}{l}{$^{**}$\footnotesize{This parameter is used to derive the initial mass of gas (see equation~\ref{eq_gas_f_i}).}} \\
\label{tab_param}
\end{tabular}
\end{table*}

We used the mean and the standard deviation of the fitted functions to set the typical
value and the uncertainty of our model input parameters (Table~\ref{tab_param}).
However, it should be kept in mind that these numbers are the result
of assuming that all the works presented in this section are both
independent and equally significant.  Neither of these assumptions is
true, strictly speaking; multiple different observational estimates
use the same astronomical objects to derive constraints, and the
techniques used are of varying quality and sophistication, and have
different attendant systematic errors.  That said, the degree of
accuracy of the assumptions of independence and equal significance is
both hard to quantify and to express statistically, and thus we use
the fitting technique described above to give an approximate
description of the uncertainty in our knowledge.

To incorporate these input uncertainties into galactic chemical evolution
predictions, we use a Monte Carlo algorithm and run several simulations where the value of each
parameter are assumed to be independent, and are randomly selected
using its fitted Gaussian function as a PDF. 
When using this approach, for a parameter having a PDF given by $g(x)$, the probability
of randomly selecting a certain value $x_r$ is directly proportional to $g(x_r)$.
In other words, values close to the mean value $x_0$ are more likely to be
picked than values that are further away, simply because the Gaussian PDF $g(x)$ reaches its
global maxima when $x=x_0$.  As shown in Figure~\ref{fig_param}, some PDFs
can generate negative values for our parameters, which is not physical.  In 
the code, we limit our random selection to positive values only.

\begin{figure*}
\includegraphics[width=7.in]{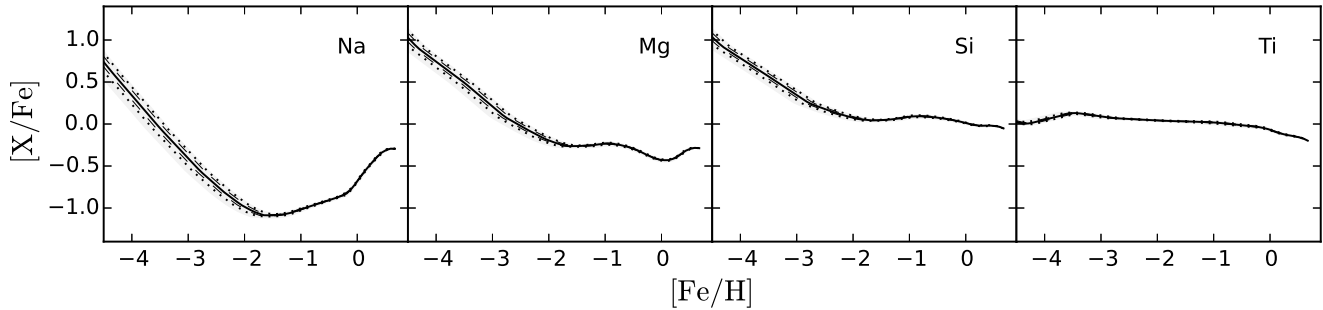}
\caption{Predicted chemical abundances relative to Fe and their uncertainties,
  generated by varying only the current stellar mass ($M_{\star,f}$) and the 
  current mass of gas ($M_{\mathrm{gas},f}$),
  as a function of [Fe/H] for four elements. The solid lines represent the median values whereas
  the dashed and dotted lines are the 68\% and 95\%
  confidence levels.}
\label{fig_chem_evol_mgal}
\end{figure*}

This method enables the exploration of a wide range of model input
parameters, and also takes into consideration the notion that some
values of these parameters are more 
probable than others.  By running a significant number of simulations
using this technique, one ends up having many predicted model outputs that cover a
range of possible solutions, but where most of the predictions
are located near the most probable solution.  In other words, the
PDF of an input parameter, as processed
by the model, induces a related distribution function of predictions related to
the evolution of each element as a function of [Fe/H].  Given the
complexity of even a simple chemical evolution model, the
distribution functions of the input parameters and the model outputs
are not necessarily functionally related -- as shown
in the next sections, even if Gaussian functions are assumed for the
PDF of each input parameter, the induced PDFs of the observational
predictions are not necessarily Gaussian.

\section{Results}
\label{sect_results}
The goal of this paper is to illustrate and quantify the impact of
input parameters on the level of uncertainty associated with one-zone galactic
chemical evolution calculations.  To do so, we first ran 700 simulations\footnote{We did a convergence test and found that
beyond 700 runs, for the number of parameters that we are using within our model, the results are converged.} where all the
input parameters were independently and simultaneously selected using the Gaussian
probability distribution functions described in Section
\ref{sect_pdf}.  Then, we took each parameter individually and ran
an additional set of 300 simulations\footnote{We found that results need less runs to converge when only
one parameter is varied.} where we only varied the
considered parameter and kept all the others at their most probable value.
This enabled us to have an idea of the contribution of each parameter
on the overall uncertainty.  We also ran another set of 300 simulations
where we simultaneously varied $M_{\star,f}$ and $M_{\mathrm{gas},f}$
(see Section~\ref{sect_speed_ece}), since they both affect our numerical 
predictions in a similar way.

Throughout this section, we do not show any figure regarding the lower
and upper mass limits of the IMF, because our simulations demonstrated
that these two parameter do not generate a significant amount of
uncertainty.  This is mainly because the stellar yields in this work
are only applied on stars with initial mass between 1 and $30\,$M$_\odot$
(see Sections~\ref{sect_m_up} and \ref{sect_y_uml} for discussions).

\subsection{Speed of the Early Chemical Enrichment}
\label{sect_speed_ece}

Although the current analysis is very specific to our choice of
galactic setup, this section gives the logic behind the generation of
uncertainty caused by the parameters regulating the final mass of gas and the
stellar mass formed during a simulation.  Figure~\ref{fig_chem_evol_mgal} presents chemical
evolution predictions for four elements where only $M_{\star,f}$ and $M_{\mathrm{gas},f}$
have been varied.  For each plotted element, the gray shaded
area illustrates at a given metallicity the full range of possible solutions, whereas the
solid, dashed, and dotted lines represent the median
value and the 68\% and 95\% confidence levels, respectively.  As shown in this
figure, there is a clear correlation between the level of
uncertainty and the steepness of the slope of [X/Fe] for [Fe/H]~$\lesssim$ $-2$.

\begin{figure*}
\includegraphics[width=7in]{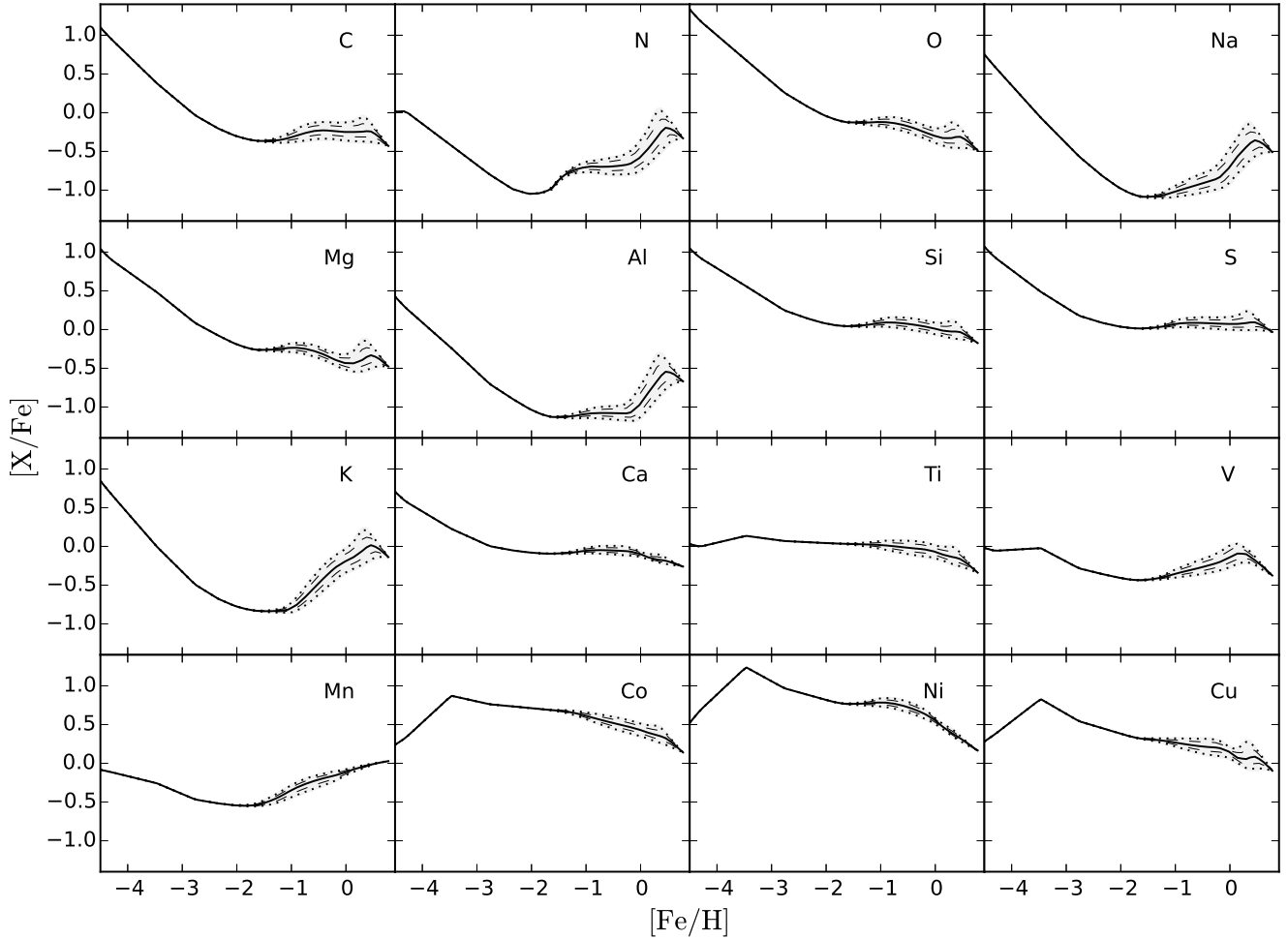}
\caption{Predicted chemical abundances relative to Fe and their uncertainties,
  generated by varying only the number of SNe Ia ($N_\mathrm{Ia}$),
  as a function of [Fe/H] for 16 elements.  The lines are the same as in Figure
  \ref{fig_chem_evol_mgal}.}
\label{fig_chem_evol_sn1a}
\end{figure*}

Modifying the final mass of gas and the final stellar mass varies the initial stellar-to-gas
mass ratio at the beginning of our simulations.  The early [Fe/H] concentration is therefore
affected by these parameters because the stellar mass sets the amount of iron ejected by stars and
the mass of gas sets the amount of hydrogen in the gas reservoir.  However, having a larger
or smaller stellar concentration does not modify the [X/Fe] abundances,
since those are determined by the ejecta of the first stellar populations.  The uncertainties seen in
Figure~\ref{fig_chem_evol_mgal} are then mainly generated by
horizontal shifts toward higher or lower [Fe/H], which have a greater impact
when the element under consideration has a steep slope in its [X/Fe] evolution.
However, overall, the uncertainties generated by $M_{\star,f}$ and $M_{\mathrm{gas},f}$
are not significant in our case.

\subsection{SNe Ia and Late Enrichment}
\label{sect_results_Ia}
Figure~\ref{fig_chem_evol_sn1a} presents the uncertainty caused by
$N_\mathrm{Ia}$, the total number of Type Ia explosions that occur per
unit of solar mass formed in an SSP.  The first
noticeable feature is the lack of uncertainty below
[Fe/H]~$\sim$~$-1.5$, which is caused by the delay between the
formation of intermediate-mass stars and the onset of the first SNe~Ia.
In our case, this [Fe/H] value is associated with a galactic age
of $\sim\,150$~Myr.
For the Milky Way, observations
suggest that SNe~Ia only started to be relevant around [Fe/H]~$\sim$~$1$ (e.g., \citealt{mg86,cmr01}).
This discrepancy is mainly caused by our closed-box assumption.
Indeed, by having the entire gas reservoir at the beginning of 
simulations, instead of gradually adding gas with inflows, there is an undesirable high initial 
gas-to-stellar mass ratio that dilutes the metals ejected
by the first stellar populations (see also \citealt{lb75,c80}).  This reduces the rate at which [Fe/H]
is increasing and moves the onset of SNe~Ia to metallicities
lower than what is expected from observations.

\begin{figure*}
\includegraphics[width=7in]{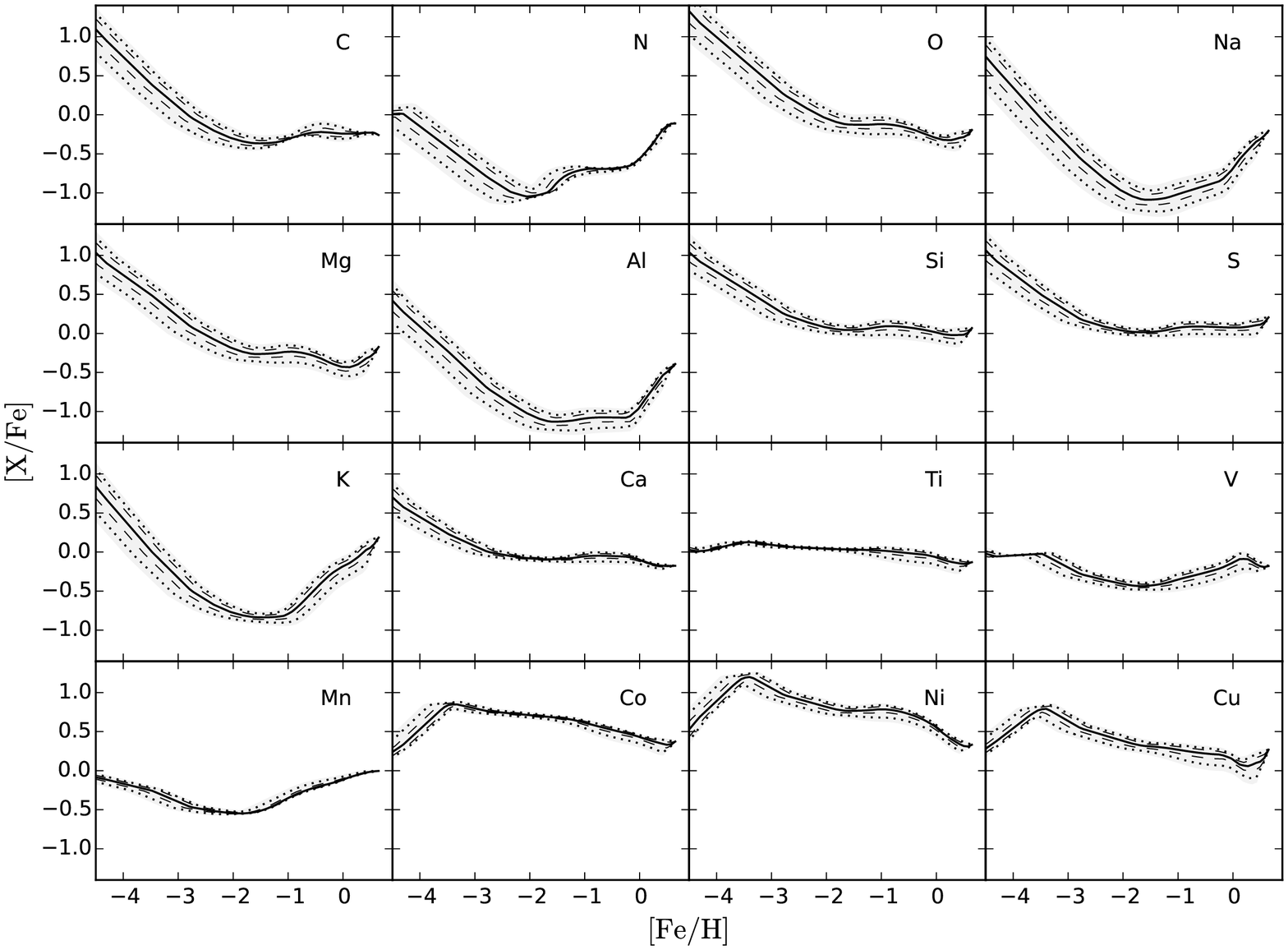}
\caption{Predicted chemical abundances relative to Fe and their uncertainties,
  generated by varying only the slope of the high-mass end of the
  stellar IMF ($\alpha$), as a function of
  [Fe/H] for 16 elements.  The lines are the same as in Figure
  \ref{fig_chem_evol_mgal}.}
\label{fig_chem_evol_alpha}
\end{figure*}

For the vast majority of elements presented in Figure~\ref{fig_chem_evol_sn1a}, modifying the number of SNe Ia
generates a diagonal shift in the predictions (along the upper left to
lower right diagonal).  For example, adding
more Type Ia explosions will increase [Fe/H] but will also decrease
[X/Fe].  This is why there is generally a diagonal cut in the
uncertainty shape at high [Fe/H].  For metallicities that are below
the beginning of this diagonal cut, all the 300 possible predictions
are included in the statistical analysis.  But at higher
metallicities, only a fraction of these predictions have sufficient SNe~Ia
to reach such high iron abundances.

As seen in Figure~\ref{fig_chem_evol_sn1a}, the level of uncertainty
is not always the same from one element to another.  Elements that are
not significantly produced by SNe Ia, such as the CNO elements, are
the most affected by the $N_\mathrm{Ia}$ parameter.  This is because
the variation of the number of explosions only modifies the iron
content in the [X/Fe] vs [Fe/H] relations.  But in the case of other
elements, such as Ca and Ni, for which SNe Ia have a contribution
of about 10\% to 50\% (according to our set of yields), the predictions become less affected by the
number of explosions.  The idea behind this trend is that increasing
$N_\mathrm{Ia}$ \textit{forces} the gas reservoir to look like the
ejecta of an SN Ia.  Since these explosions already contribute
significantly to the production of such elements, the chemical
abundances relative to iron eventually become saturated when $N_\mathrm{Ia}$ is
increased.

Because of the direction of the diagonal shift in the [X/Fe] vs [Fe/H]
space caused by the variation of $N_\mathrm{Ia}$, elements showing a
positive slope in their evolution (increasing [X/Fe] with increasing [Fe/H]) will systematically tend to have a
higher level of uncertainty than the ones showing a negative slope
(decreasing [X/Fe] with increasing [Fe/H]).  Mn is an exception, since
this element and Fe are both mainly ejected by SNe~Ia.  As seen in \cite{tht09}, the evolution of alpha elements relative to
iron in nearby dwarf spheroidals usually shows a steeper negative slope
than in the Milky Way.  This means that the variation of
$N_\mathrm{Ia}$ should generate less uncertainty when tuning a
galactic chemical evolution model to represent a dwarf galaxy, even if
SNe Ia contribute significantly to their chemical evolution
(e.g., \citealt{v12}).  As was pointed out in the previous section,
the level of uncertainty is tightly connected to the specific shape
and slope of each prediction, which in turn is directly related to
the choice of stellar yields and the type of galaxy considered.

\begin{figure*}
\includegraphics[width=7.0in]{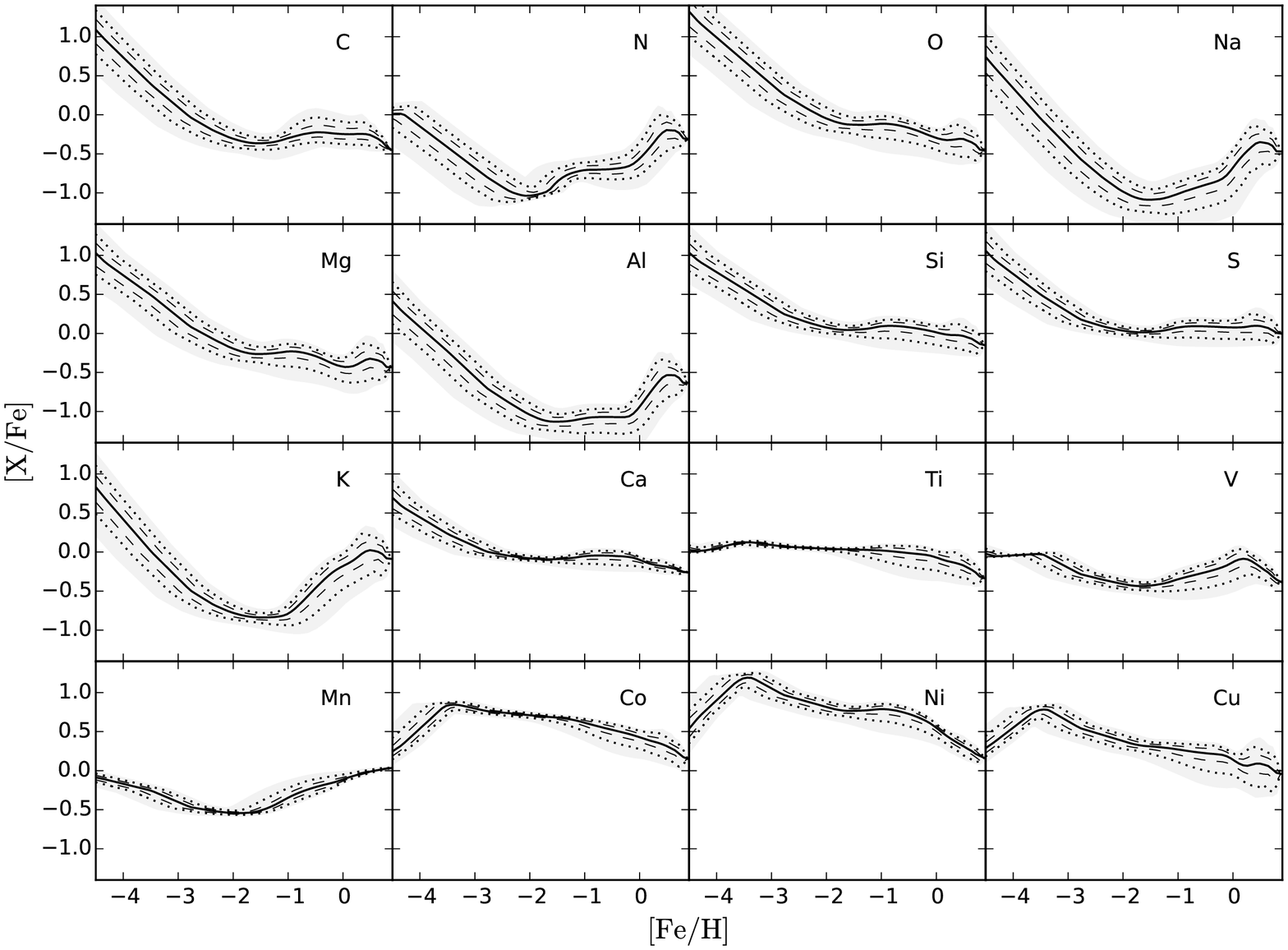}
\caption{Predicted chemical abundances relative to Fe and their uncertainties as a
  function of [Fe/H] for 16 elements, including all model parameters
  varied simultaneously.  The lines are the same as in
  Figure~\ref{fig_chem_evol_mgal}.}
\label{fig_chem_evol}
\end{figure*}

The slope of the DTD function of SNe~Ia, $\beta$, 
has an impact similar to that of the total number of explosions in the sense that it also produces
diagonal shifts in the predictions.  However, the slope of
the DTD function does not modify the total number of explosions
associated with each stellar population.  The uncertainties generated by
$\beta$ are on average three times lower than the ones generated by
$N_\mathrm{Ia}$, which is mainly caused by the different input level
of uncertainty of these two parameters (see Table \ref{tab_param}).

\begin{figure*}
\includegraphics[width=7in]{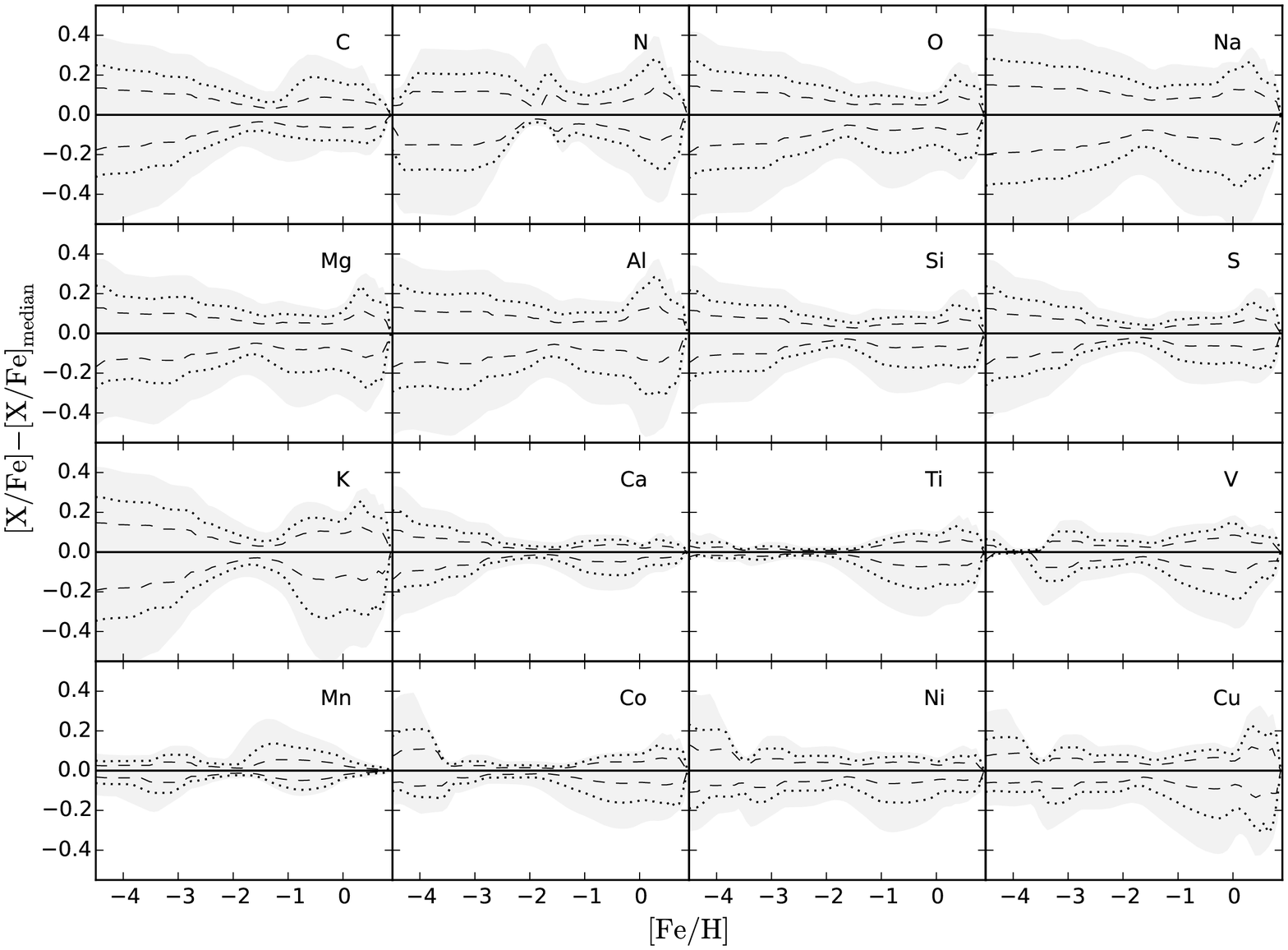}
\caption{Uncertainties in our predictions as a
  function of [Fe/H] for 16 elements, including all model parameters
  varied simultaneously.  This is the analogous of Figure~\ref{fig_chem_evol}
  where the median values have been subtracted from all predictions.}
\label{fig_chem_evol_delta}
\end{figure*}

\subsection{Slope of the IMF and Stellar Yields}
\label{sect_results_imf}
The slope of the high-mass end of the IMF is the most
significant parameter in the generation of uncertainties in our model.
As seen in Figure~\ref{fig_chem_evol_alpha}, this parameter impacts
the predictions at every metallicity.  Although the uncertainties vary
substantially from one element to another, the most important notion to
remember is that the IMF mostly affects the prediction when iron and the
considered elements are produced in different stars.  Otherwise,
modifying the relative numbers of stars with different initial mass
will not change the [X/Fe] abundance ratios, although it can modify
[Fe/H].

It is instructive to consider the very-low-metallicity parts of Figure
\ref{fig_chem_evol_alpha} below [Fe/H]~$\sim$~$-2$.  At these
metallicities, no SN Ia has occurred yet and the chemical evolution is
then purely driven by the yields of massive stars at Z~$=$ 0 and
10$^{-4}$ (see Section \ref{sect_trans_Z}), which simplifies the analysis.  As an example, in the case of C, N, O, Na,
Mg, and Al, the level of uncertainty has a
continuous-looking shape, since these elements and Fe are not ejected
by the same stars, both at Z~$=$ 0 and 10$^{-4}$.
For Si, S, K, and Ca, there is a tightening feature around [Fe/H]
of $-2$ since these elements and Fe are mainly ejected by
the same stars at Z~$=$~10$^{-4}$, but by different stars at Z~$=$~0.
Only the very-low-metallicity end of [X/Fe] is then affected by
the IMF.  For Ti and Mn, their evolution do not significantly depend on the slope of the IMF since
they are all generally ejected by the same stars that the ones responsible for
the ejection of Fe, for Z~$=$ 0 and 10$^{-4}$.  
This demonstrates that the uncertainties
produced by the slope of the IMF are directly connected to the choice
of stellar yields and the number of metallicities available with these yields.
The same logic is applicable at higher
metallicities.  The analysis is, however, somewhat more complex
since iron now comes from both the CC~SNe of young stellar populations
and the SNe Ia of older stellar populations.

\begin{figure*}
\includegraphics[width=7in]{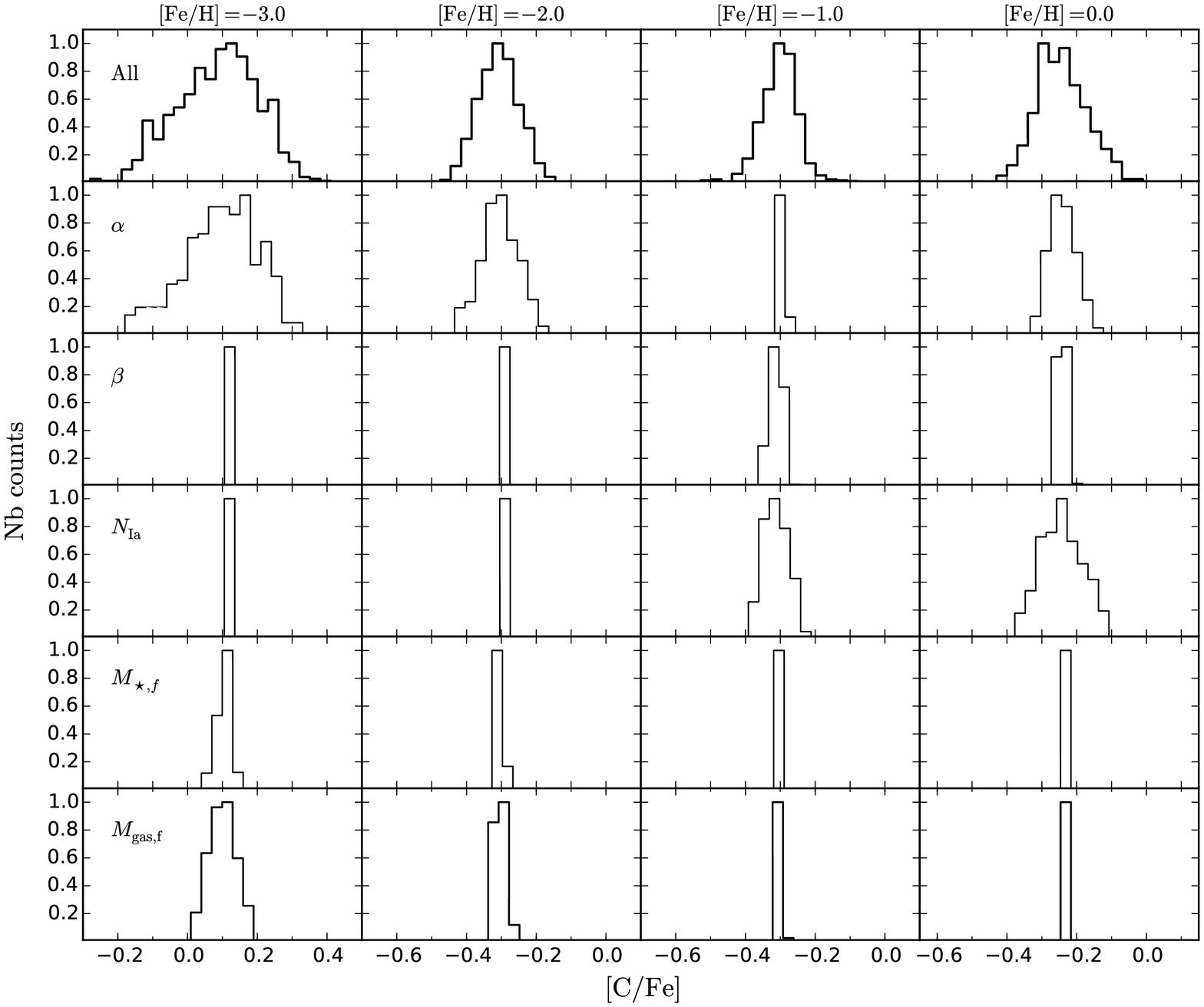}
\caption{Probability distribution functions of [C$/$Fe] taken at different [Fe$/$H].  In the first row, all the parameters are included in the uncertainties.  In the remaining rows, each parameter has been considered individually in order to provide an estimate of their contribution to the overall uncertainty.}
\label{fig_hist_Mg}
\end{figure*}

Carbon shows an interesting low level of uncertainties around [Fe/H]
of $-$1 (see Figure~\ref{fig_chem_evol_alpha}).  This special
case corresponds to a cross-over point, which means that a  
prediction below the median at [Fe/H]~$<$~$-$1 will be above the median
at [Fe/H]~$>$~$-$1, and vice versa.  From our set of
stellar yields, carbon is mainly ejected by low-mass stars,
which explains the [C/Fe] bump seen at [Fe/H]~$\sim$~$-0.5$.  In
addition, carbon yields at $Z=10^{-4}$ are 50\% larger compared to
yields at $Z=0.001$, and 85\% larger compared to yields at $Z=0.006$.
Having a steeper IMF creates more low-metallicity stars, since [Fe/H]
initially increases at a slower pace due to the reduced number of
massive stars.  This case corresponds to the lines far below the
median line of [C/Fe] at early times.  However, once low-mass stars start
to contribute to the ejection of carbon, the high number of
low-metallicity stars maximizes the amplitude of the bump located at
[Fe/H]~$\sim$~$-0.5$, thus creating the cross-over point.
\\
\\
\subsection{Overall Uncertainty}
\label{sect_ov_un}
Figures \ref{fig_chem_evol} and \ref{fig_chem_evol_delta} show the overall uncertainty associated
with our predictions coming from our one-zone model, which is the
result of varying all the parameters simultaneously.  The overall
level of uncertainty evolves as a function of [Fe/H] and 
varies from one element to another.  By combining these figures
with the results presented in the previous sections, it is now possible to
analyze the uncertainty of each element in a different manner.
As an example, the top row of Figure~\ref{fig_hist_Mg} presents the
overall PDFs of [C/Fe] for four
different [Fe/H] values, whereas the following rows illustrate the
contribution of the individual parameters.  This figure is
useful for visualizing which input parameters have the most impact and at
which metallicity.  In the case of carbon, the slope of the IMF is the main contributor of the overall
uncertainties at low [Fe/H] (i.e., early in the galaxy's star
formation process).  At high [Fe/H] (i.e., later in time), the total
number of SNe Ia dominates the uncertainties.

\begin{figure*}
\includegraphics[width=7in]{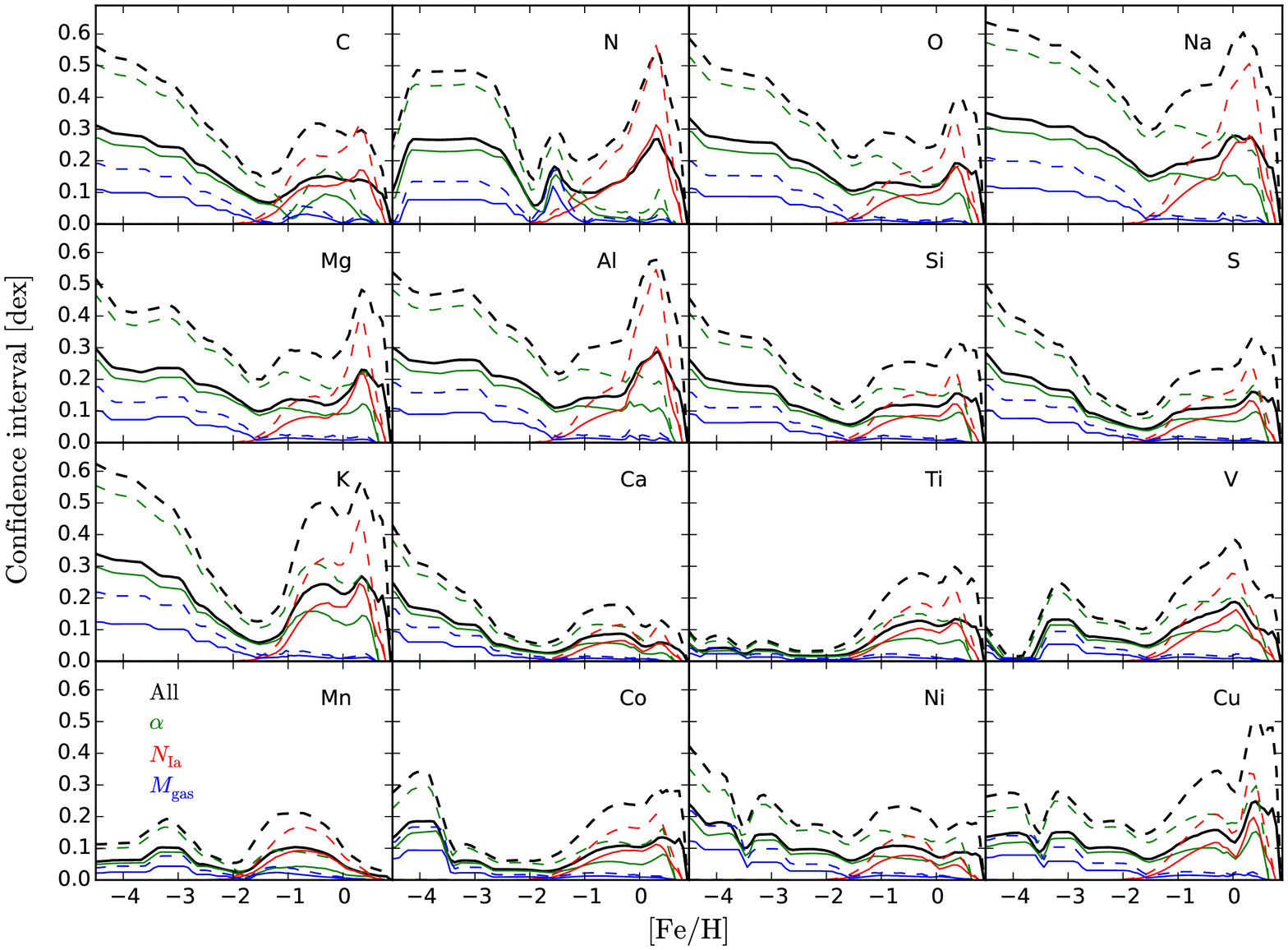}
\caption{Evolution of the confidence intervals of [X/Fe] as a function of [Fe/H]
  and parameters.  The solid and dashed lines represent respectively
  the thickness of the confidence intervals within 68\% and 95\%.  The
  black lines include all seven parameters in the generation of the
  uncertainties.  The coloured lines, as indicated, show the estimated
  individual contribution of the three most important parameters.}
\label{fig_ds}
\end{figure*}

It is worth noting that summing all of the uncertainties obtained by
varying the individual parameters one at a time will not reproduce
the overall uncertainty presented in Figures \ref{fig_chem_evol} and \ref{fig_chem_evol_delta}.
Indeed, when varying only one parameter, its uncertainty is only 
induced into a specific prediction, since all of the other parameters are kept
constant.  However, when all parameters are varied at the same time,
the uncertainty induced by a single parameter becomes spread in all
the possible predictions generated by the random selection of the other
parameters.  It is then important to reiterate that our analysis of
individual \textit{contributions} is only an estimate.  Nevertheless,
it turns out to be an efficient way to evaluate which parameter
generates the most uncertainties in galactic chemical evolution
models.

In Figure~\ref{fig_ds}, we show a different representation of the
information contained in Figure~\ref{fig_hist_Mg}, but for all 16
elements discussed in prior sections.  We calculate the thickness of the
68\% and 95\% confidence intervals for every element as a function of [Fe/H].
This figure only includes the overall uncertainties
and the contributions of the three parameters that contribute the most
to the global uncertainty, which
are the slope of the stellar IMF, the total number of SNe~Ia, and the final
mass of the gas reservoir of the galaxy.  A local minimum in the overall uncertainty is often seen
around [Fe/H]~$\sim$~$-1.5$, which is partially caused by the rise of
SNe~Ia.  For some elements such as Ti, the contribution of the
IMF also starts to increase at the same metallicity.   As
mentioned throughout this paper, the derived amount of uncertainty also
depends on the stellar yields and should be taken with caution.

\subsection{Impact of the Reference Element}
\begin{figure}
\includegraphics[width=3.25in]{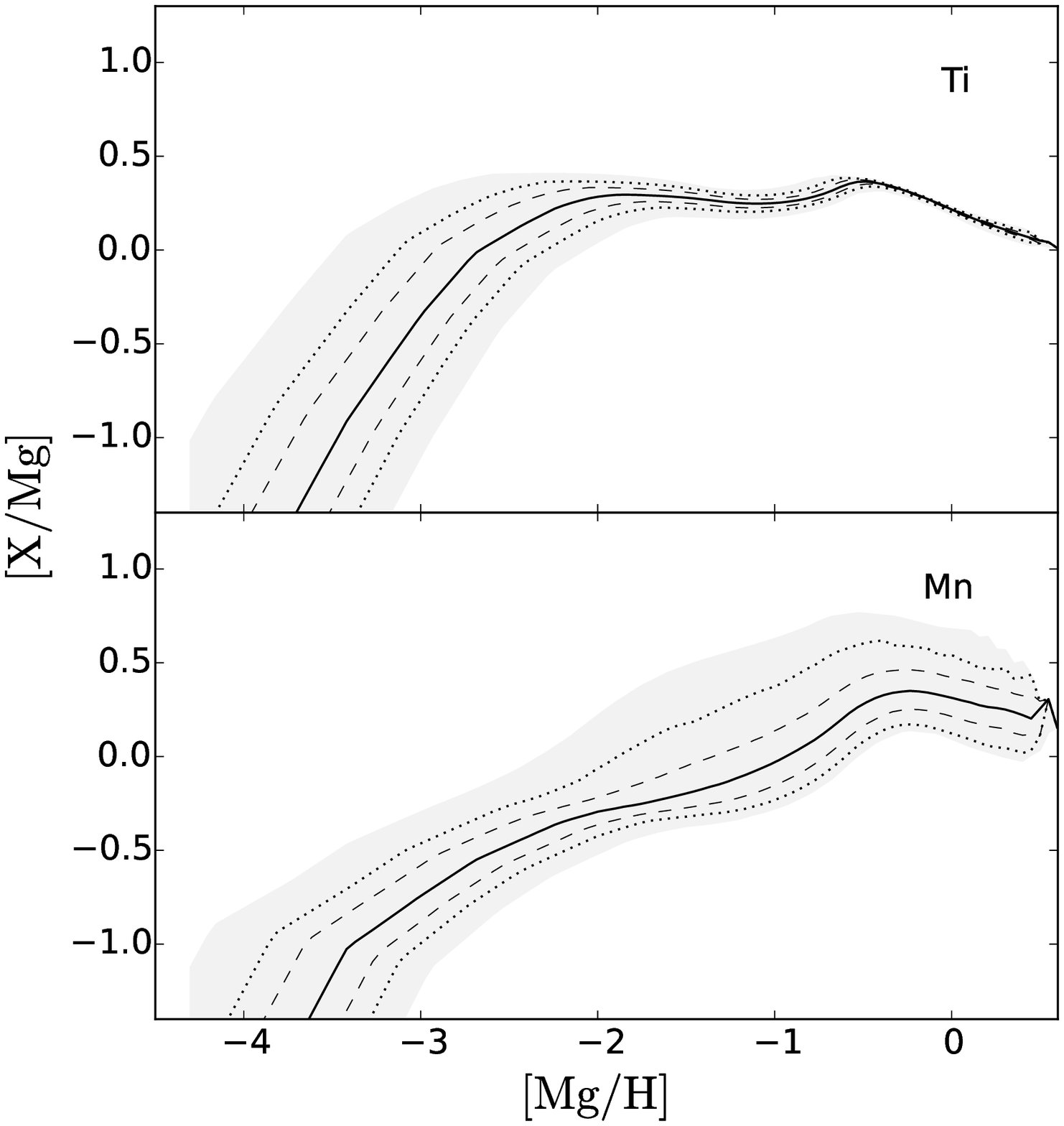}
\caption{Predicted chemical abundances and their
  uncertainties as a function of [Mg/H] for Ti and Mn,
  including all model parameters varied simultaneously.
  The lines are the same as in Figure~\ref{fig_chem_evol_mgal}.}
\label{fig_Mg_H}
\end{figure}

Although the measurement of Fe has
observational conveniences, other elements can be used 
to study the chemical evolution of galaxies (see the discussion found in \citealt{c04}).
As an example, the upper panel of Figure~\ref{fig_Mg_H} presents the overall
uncertainty of [Ti/Mg] as a function of [Mg/H].
Compared to Figure~\ref{fig_chem_evol}, the uncertainties
in the predictions of Ti at high metallicity are greatly reduced when
plotted relative to Mg instead of Fe.  This is partially because SNe~Ia
are not the main source of production of Mg and Ti, which
eliminates large amount of uncertainty generated by $N_\mathrm{Ia}$,
the number of SNe Ia in SSPs (see Section \ref{sect_results_Ia}).  On the other hand,
because of the uncertainty in the slope of the IMF, the predictions are a lot
more uncertain at low metallicity when using Mg instead of Fe, since 
Ti and the reference element (Mg) are no longer ejected by the same
stars in our stellar yields at $Z=0$ and $Z=10^{-4}$.

As seen in the lower panel of Figure~\ref{fig_Mg_H}, the
predicted evolution of Mn is now very uncertain when plotted
relative to Mg, despite the fact that Mn was one of the elements with the
lowest level of uncertainty when plotted relative to Fe (see Figure
\ref{fig_chem_evol}).  The element of reference then have a
significant impact on the amount of uncertainties generated
by input parameters.  As a general 
remark, the predicted evolution of [X/X$_\mathrm{ref}$] as a
function of [X$_\mathrm{ref}$/H] will always be more uncertain
when the elements X and X$_\mathrm{ref}$ are not ejected by
the same stars.
\\

\subsection{Evolution of Individual Elements}
\begin{figure*}
\includegraphics[width=7in]{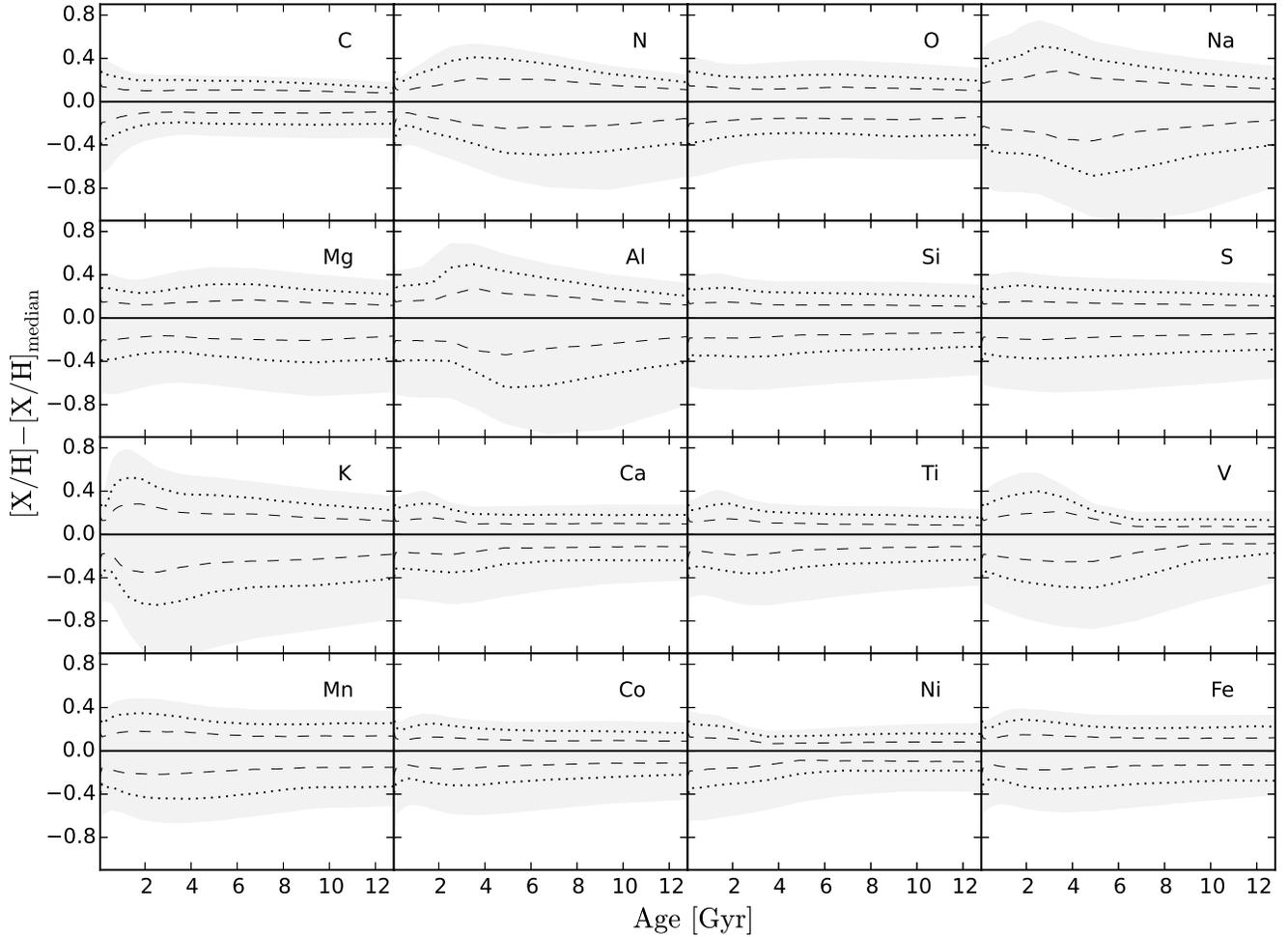}
\caption{Uncertainties in our predictions as a function of galactic age for 16 elements relative to H, including all model parameters varied simultaneously. The lines are the same as in Figure~\ref{fig_chem_evol_mgal}.}
\label{fig_X_H_vs_t}
\end{figure*}

Looking at the abundances of one element against another one ([X/X$_\mathrm{ref}$] vs [Y/X$_\mathrm{ref}$])
is convenient for comparing with observational data.  However, it is hiding the uncertainties in the evolution 
of individual elements.  As shown in Figure~\ref{fig_X_H_vs_t}, the amount of uncertainty is larger for
[X/H] as a function of time than for [X/Fe] as a function of [Fe/H] (see Figure~\ref{fig_chem_evol_delta}).
This means the overall amount of uncertainty associated with one specific element depends
on the way numerical predictions are shown.  For example, Figure~\ref{fig_chem_evol_delta} suggests
that our predictions for Ti are almost free of uncertainty at low metallicity, which is a misleading 
conclusion when referring to Figures~\ref{fig_Mg_H} and \ref{fig_X_H_vs_t}.  
In this last figure, there is a certain similarity in the shape of the uncertainty envelope (gray shaded area) for
Na, Al, and K.  As seen in Figure~\ref{fig_chem_evol}, these three elements show a similar \textit{U}-shape in the [X/Fe]-[Fe/H] space, reinforcing the idea that
the level of uncertainty also depends on the global shape of numerical predictions.

\section{Discussion}
\label{sect_disc}

So far, we have demonstrated that some input parameters in 
a closed-box, single-zone model can induce significant amount of
uncertainties in numerical predictions.  This is an important step in
our long-term project which aims to better quantify the uncertainties
inherent in more complex chemical evolution models.
Our study is, however, subject to a variety of limitations and caveats.
In the next sections, we briefly discuss some of them and highlight
additional sources of uncertainty in chemical evolution studies.

\subsection{Stellar Yields}
\label{disc_yields}
For the sake of simplicity, we choose to use a single set of yields,
which includes five metallicities between $Z=10^{-4}$ and $Z=0.02$
for massive and AGB stars (see Section \ref{sect_stel_yie}), the SN~Ia
yields of \cite{tny86}, and the zero-metallicity yields of \cite{hw10} for the
first stellar populations formed in our simulation.  It is known that
different stellar evolution codes and different research groups
produce different nucleosynthetic yields (\citealt{g97,g02,rktm10,mcgg15}).  
But, besides the SN~Ia and zero-metallicity yields, all of our stellar
models have been calculated with the MESA code and were post-processed
with NuGrid's nucleosynthesis code MPPNP in order to have
a consistent set of assumptions.
  
\begin{figure}
\includegraphics[width=3.25in]{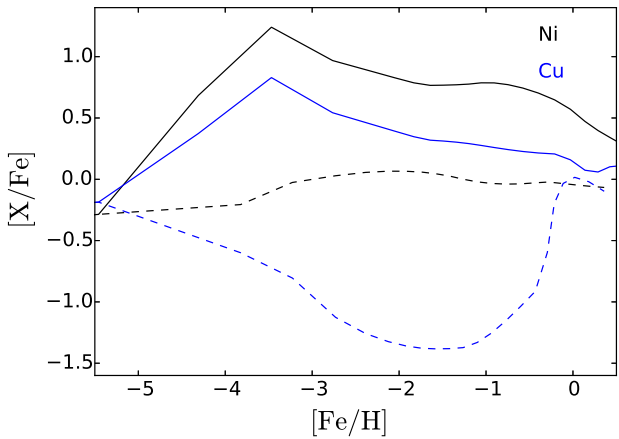}
\caption{Predicted abundances of Ni and Cu, relative to Fe, as a function of [Fe/H] using two different mass-cut prescriptions for the CC~SN yield calculations.  The solid lines represent the prescription of Fryer et al.~(\citeyear{fbw12}), designed to reproduce the observed neutron-star and black-hole mass distribution functions, while the dashed lines represent the \textit{Ye} prescription (see text in Section~\ref{disc_yields}).  For this figure, we set all parameters to their typical value (see Table~\ref{tab_param}).}
\label{fig_Ni_Cu}
\end{figure}

However, even if a set of yields
is calculated with the same code, there are always sources of uncertainties
attached to nuclear reaction networks and modeling assumptions.
For example, the choice of mass cut which defines where 
the explosion is launched inside a massive star has a significant
impact on chemical evolution predictions.  In this work, we used a prescription derived 
by \cite{fbw12} (see also \citealt{p13a}) to tune the mass cut of our stellar models in order
to reproduce the observed neutron-star and black hole mass distribution
functions (\citealt{bbf12}).  The explosive yields generated
by this prescription are different than those generated with 
a mass-cut prescription based on the electronic mass fraction value ($Y_e$).
This is particularly true for Ni and Cu, as shown in Figure~\ref{fig_Ni_Cu}.
For the explosive yields produced with the $Y_e$ prescription, we
launched the explosion from the location where $Y_e=0.4992$
(see \citealt{a96} for more information).
It is however beyond the scope of this paper
to fall into a detailed analysis of the impact modeling assumptions in CC~SN calculations.
A description of the yields produced by different mass-cut
prescriptions will be presented in C. Ritter et al. (in preparation).

\subsection{Interpolation of Stellar Yields}
\label{sec_int_stel_yie}
Because of the large metallicity range covered in galactic chemical evolution, yields are
usually interpolated according the logarithm of $Z$.  Although we can
interpolate between $Z=10^{-4}$ and $Z=0.02$, we cannot follow the
same interpolation law between $Z=0$ and $Z=10^{-4}$.  Thus, there is
no interpolation below [Fe/H]~$\simeq-2.3$.  The only $Z$-dependency
considered in the early chemical evolution is the transition between
the use of zero-metallicity yields and $Z=10^{-4}$ yields, which
occurs as soon as the primordial gas reservoir becomes enriched by the
first stars.  The lack of very-low-metallicity stellar yields,
especially for CC~SNe, represents an additional hidden source of
uncertainty for numerical predictions at [Fe/H]~$\lesssim$~$-2.5$.
This situation also occurs with the yields of \cite{ww95} and
\cite{n06}.  But it should be noted that efforts has been made to
provide stellar yields at very-low metallcity (e.g., \citealt{cl04,h07}).
However, the calculation of very-low-metallicity yields is challenging and
involves physical problems, such as hydrogen ingestion events, that are
difficult to solve in one-dimensional calculations (e.g., \citealt{herwig11}).

Even when interpolation is possible between metallicities, the number of metallicity bins in
our input yields, and also the number of stellar mass bins, may substantially
affect our results.  The extent of this is unclear -- no study has been made
to determine the ``minimum acceptable'' grid of models.

\subsection{Galactic Chemical Evolution Model}
We use a closed-box model with a constant star formation
efficiency.  We explicitly ignore inflow of low-metallicity gas (i.e., accretion),
outflow of high-metallicity gas (i.e., galactic winds), and all
aspects of galaxy formation including the growth of the galactic
potential well, and stochasticity due to
mergers.  This was done deliberately -- we want to consider the
simplest possible model.  Considering more complex models
surely add more realism and flexibility in the predictions, but it also 
introduce new sources of uncertainty as more parameters and
more assumptions are implemented.  In future work, we plan to highlight
this by expanding our uncertainty study beyond the closed-box
model.

\subsection{Initial Mass Function}
Although we used a Chabrier IMF, there are other
possible IMFs (Kroupa, Salpeter, lognormal, etc.) that have been
used to fit the local observational data.  We cannot readily
include them simultaneously with the approach described in Section~\ref{sect_pdf},
because these IMFs have different functional forms
and often possess discontinuities.  We can perform our analysis for
other IMFs, but the results are likely to remain qualitatively the
same, though quantitatively somewhat different.
We further assume that the stellar IMF is the same in
  all young star clusters, and implicitly everywhere in the universe
at all times.  For example, we assume extremely low-metallicity stars have
the same IMF as solar-metallicity stars.  This is not particularly
defensible -- environment and redshift possibly have an effect on stellar IMF
(e.g., \citealt{t07,capp12,cvd12,c15}).
The PDF of our input parameters defining the IMF could in fact
vary as a function of time in our simulations,
which would modify the contribution of $\alpha$, $M_\mathrm{low}$, 
and $M_\mathrm{up}$ on the generation of uncertainties.

We only include the contribution
of CC~SNe for stars with initial mass below $M_\mathrm{thresh}=30\,$M$_\odot$.  If 
we were to include them, deriving the yields of those
more massive stars becomes tricky and uncertain, since we need to rely on an extrapolation rather than
on a safer interpolation between stellar models.  The result of such
an extrapolation depends on the available maximum stellar initial mass included
in the set of yields, which is $25\,$M$_\odot$ in our case (except for the
zero-metallicity yields of \citealt{hw10}).  Because of this, and because $M_\mathrm{thresh}$
is not yet an observationally estimated quantity, addressing
this issue in more detail is beyond the scope of this paper.  Moreover,
there are still uncertainties regarding the fate of stars more massive than $\sim$~$15\,$M$_\odot$
(see Section~\ref{sect_y_uml}).  

\subsection{Generation of Uncertainties}
The method that we use to calculate the Gaussian distributions
  describing our input parameters is relatively simple -- we create a
  distribution from the normalized values of the parameters from a
  variety of observations and fit a Gaussian distribution to it, which
  gives us a mean and a standard deviation.  This assumes that the observations
  are independent and use the same methods, which is
  not true.  In reality, most observational works use different approaches varying in accuracy and sophistication.
  It is difficult and typically highly subjective to quantify the statistical weight of
  each observed quantity, which would nevertheless be useful in order to generate more representative PDFs.

We assume that all of our input parameters are uncorrelated.  This allows
us to randomly and independently choose their value in each simulation.  However,
there must be correlations, especially between the slope of the IMF and the number
of SNe~Ia.  Adding a correlation between these two parameters would increase
the amount of uncertainty at the high-metallicity end of our [X/Fe]-[Fe/H] predictions.
All the 16 considered elements, besides C, N, and Mn, are mainly
ejected by massive stars.  A correlation between the IMF and the number of SNe~Ia
would therefore favors more SNe~Ia (more iron) when the number of massive
stars is reduced, and vice-versa.  For C and N, such a correlation would reduce
the level of uncertainty, as these elements are mainly ejected by low- and intermediate-mass
stars, which includes the progenitors of SNe~Ia.

Some parameters have not been included in the
generation of the uncertainties (see Section~\ref{sect_other_param}).  In addition, we did not considered
uncertainties associated with stellar yields, modeling assumptions,
star formation histories, and galaxy evolution environment.
The uncertainties derived in this paper thus represent a lower limit
of the real uncertainties inherent in chemical evolution models.

\section{Summary and Conclusions}
\label{sect_conc}
Using a simple chemical evolution code, the goal of
this paper was to quantify the uncertainties induced in numerical predictions
by seven input parameters describing the IMF, the rate of SNe~Ia, and
the basic properties of the Milky Way.  To do so, we compiled observational and
numerical work from the literature in order to derive the typical value and the
PDF of each parameter.  We then ran several hundred simulations using
a Monte Carlo algorithm to randomly select the input parameters from their
individually determined PDFs.  Using this approach, we have been able to quantify in a statistical manner
the uncertainties in the chemical evolution of our simulated galaxies
by providing the most probable predictions along with their 68\% and 95\%
confidence levels.  Our main conclusions are the following.

\begin{itemize}

\item When considering the variation of all parameters simultaneously,
these parameters produce an overall uncertainty roughly between
0 and 0.6 dex when abundances are plotted against metallicity.
However, the level of uncertainty is
metallicity-dependent and is different from one element to
another, since every element has its own evolution pattern and
production site.  The amount of uncertainty can reach 1 dex
when looking at the evolution of an individual element as a function
of time instead of metallicity.

\item The overall uncertainties are
produced by a combination of different input parameters that
contribute at different metallicities.  The
current mass of gas and the current stellar mass of the galaxy only affect
the early chemical evolution below [Fe/H]~$\sim$~$-2$.  On the other
hand, the slope of the delay-time distribution function and the total number of SNe~Ia only
affect the predictions above [Fe/H]~$\sim$~$-1.5$, whereas the slope of
the IMF has generally an impact at all metallicities.

\item Among the seven input parameters included in this
study, the slope of the high-mass end of the stellar IMF and the total number of SNe~Ia
contribute the most to the overall uncertainties when abundances
are plotted against metallicity.  These parameters
have a stronger impact when the considered element and the reference
element (e.g., iron) are not ejected by the same stars.

\item Input parameters do not modify the overall trends seen in numerical predictions.
Characteristic features in the [X/Fe]$-$[Fe/H] or [X/Mg]$-$[Mg/H] space,
such as the slope and the curvature of the predictions, or the presence of bumps,
cannot be modified by varying the input parameters.  Such features are 
mainly caused by the choice of stellar yields and the type of galaxy considered.
In other words, the input parameters only spread their uncertainties on top of 
predictions already \textit{pre-defined} by these two last ingredients.

\end{itemize}

Our work showed that some input parameters in one-zone, closed-box
calculations add a significant amount of uncertainty that is
often comparable to or even larger than the scatter seen in observational data, despite the
fact that we did not include the uncertainties associated with stellar yields,
galaxy formation, and modeling assumptions.  It is then very
clear that the uncertainties derived in this paper represent
only a lower limit of what must be the real and \textit{concerning}
amount of uncertainty inherent in galactic chemical evolution models.
For the moment, we cannot apply our uncertainty quantification
to more realistic models such as multi-zone and open-box models or
chemo-dynamical simulations, since further investigation is needed.
However, all of those more sophisticated models should also be affected
by their own parameters in a similar way.  Most of the parameters explored in this work, especially
for the IMF and SNe~Ia, are fundamental and must somehow be
implemented in every chemical evolution model or simulation.

\section*{acknowledgments}

This research is supported by the National Science Foundation (USA) under
grant No. PHY-1430152 (JINA Center for the Evolution of the Elements),
and by the FRQNT (Quebec, Canada) postdoctoral fellowship program.  B.W.O. was supported
by the National Aeronautics and Space Administration (USA) through grant
NNX12AC98G and Hubble Theory grant HST-AR-13261.01-A.  He was also
supported in part by the sabbatical visitor program at the Michigan
Institute for Research in Astrophysics (MIRA) at the University of
Michigan in Ann Arbor, and gratefully acknowledges their hospitality.
F.H. acknowledges support through an NSERC Discovery grant (Canada).
The NuGrid WENDI platform has been developed as part of the CANFAR
NEP project Software-as-a-service for Big Data Analysis funded by Canarie.
M.P. was supported by the Lendulet-2014 program from the Hungarian
Academy of Science (Hungary) and the Swiss National Science Foundation
(Switzerland).  S.J. is a fellow of the Alexander von Humboldt Foundation.
C.F. was funded in part under the auspices of the U.S. Department of Energy,
and supported by its contract W-7405-ENG-36 to Los Alamos National Laboratory.

\label{lastpage}

\end{document}